\newcommand{\QED}{$\Box$}

\documentclass[final]{siamltex}
\usepackage{latexsym}

\input{epsf}

\title{ The Complexity of Planar Counting Problems}

\author{
Harry B. Hunt III\footnotemark[2]
\and Madhav V. Marathe\footnotemark[3] 
\and Venkatesh Radhakrishnan\footnotemark[4] 
\and Richard E. Stearns\footnotemark[2]
}

\begin{document}

\maketitle

\renewcommand{\thefootnote}{\fnsymbol{footnote}}

\footnotetext[2]{
Email:{\tt \{hunt,res\}@cs.albany.edu}. 
Department of Computer Science, University at Albany, SUNY, Albany, NY 12222.  
Supported by NSF Grants  CCR 89-03319, 
CCR 94-06611 and CCR 90-06396.} 
\footnotetext[3]{Email: {\tt madhav@c3.lanl.gov}.
Los Alamos National Laboratory, P.O. Box 1663, MS B265, Los Alamos, NM
  87545.   Research supported by the
  Department of Energy under Contract W-7405-ENG-36.
Part of the work was done while the author was at University at Albany
and supported by NSF Grants CCR 89-03319, and  CCR 94-06611.}
\footnotetext[4]{Email: {\tt rven@cup.hp.com}. 
Mailstop 47LA-2, Hewlett-Packard Company, 19447 Pruneridge Avenue,
Cupertino, California 95014-9913. 
Part of the work was done while the author was at University at Albany
and supported by NSF Grants CCR 89-03319, and  CCR 94-06611.}

\renewcommand{\thefootnote}{\arabic{footnote}}

\begin{abstract}
We prove the \#P-hardness of the counting problems associated with various 
satisfiability, graph and combinatorial problems, when 
restricted to planar instances. 
These  problems include
\begin{romannum}
\item[{}]
{\sc 3Sat, 1-3Sat, 1-Ex3Sat, Minimum Vertex Cover, Minimum Dominating Set,
Minimum Feedback Vertex Set,  X3C, Partition Into Triangles, 
and Clique Cover.} 
\end{romannum}
We also prove the {\sf NP}-completeness of the 
{\sc Ambiguous Satisfiability} problems
\cite{Sa80} and the {\sf D$^P$}-completeness (with respect to random polynomial
reducibility) of the unique satisfiability problems \cite{VV85} associated
with several of the above problems, when restricted to planar instances.
Previously, very few {\sf \#P}-hardness results, 
no {\sf NP}-hardness results, and no {\sf D$^P$}-completeness
results were known for counting problems, ambiguous satisfiability problems
and unique satisfiability problems, respectively,
when restricted to planar instances.

Assuming {\sf P $\neq $ NP}, one corollary
of the above results is
\begin{romannum}

\item
There are no $\epsilon$-approximation algorithms for the problems of
maximizing or minimizing 
a linear objective function subject to a planar system of
linear inequality constraints over the integers.

\end{romannum}
\end{abstract}

\begin{keywords} 
planar, {\sc 3Sat}, graphs, {\sf \#P}-complete, 
                 {\sf D$^P$}-complete, {\sf NP}-complete 
\end{keywords}

\begin{AMS}

68Q15, 68R10.
\end{AMS}

\pagestyle{myheadings}
\thispagestyle{plain}
\markboth{H.B. HUNT, III, M.V. MARATHE, V. RADHAKRISHNAN AND R.E. STEARNS}{PLANAR COUNTING}



\newtheorem{claim}[theorem]{Claim}

\section{Introduction}
A number of papers in the literature 
\cite{BW91,GJ79,SP86,PB83,Va79a}, etc.,
have  considered the complexity of counting problems, proving many
such problems to be {\sf \# P}-complete. 
Other papers have studied the complexity of Ambiguous and Unique
satisfiability problems \cite{Sa80,VV85}, proving such problems to
be {\sf NP}-hard and {\sf D$^P$}-hard\footnote{Throughout this
paper, all the {\sf D$^P$}-hardness results are with respect to random 
polynomial time reductions.} respectively.
Still other papers 
\cite{DF86,DF85,GJ79,GJS76,DL82}
have considered the complexity of decision problems when restricted to
planar instances, proving many such problems to be {\sf NP}-hard. 
In this paper, we combine these lines of research and prove for the first time
that
\begin{romannum}
\item[{}]
{\bf even when restricted to planar instances},
many counting problems remain  {\sf \#P}-complete, 
many ambiguous satisfiability problems remain 
{\sf NP}-complete and many unique satisfiability problems  
remain {\sf D$^P$}-hard.
\end{romannum}
Previously, very few {\sf \#P}-hardness results, 
no  {\sf NP}-hardness results, 
and  no \\
{\sf D$^P$}-completeness
results were known for counting problems, ambiguous satisfiability problems
and unique satisfiability problems, respectively,
when restricted to planar instances.
Such results are presented for 
several satisfiability,
graph and combinatorial problems. 
These results  show that 
planar counting, planar ambiguous satisfiability and
planar unique satisfiability problems 
are {\em as hard as} arbitrary such problems, with respect to
polynomial time and  random polynomial time reducibilities.
The results in this paper both extend the 
results in the literature and also provide
additional tools 
for proving hardness results for planar
problems for various complexity classes. 
These tools include
parsimonious and weakly parsimonious
crossover boxes, the {\sf NP}-hardness of various basic planar satisfiability
problems and the {\sf NP}-hardness of the planar 
ambiguous satisfiability problems. 
(Henceforth, we denote the restriction of a
problem $\Pi$ to planar instances by {\sc Pl-$\Pi$}.)
The particular results presented here include the following:
\begin{remunerate}
\item
The problem {\sc 3Sat} has a parsimonious planar crossover box.
Among other things, this implies that the problem 
{\sc \#Pl-3Sat} is {\sf \#P}-complete, the problem\\
{\sc Ambiguous-Pl-3Sat} is {\sf NP}-complete, and the problem 
{\sc Unique-Pl-3SAT} is {\sf D$^P$}-complete. 

\item
The problem {\sc 3Sat} is simultaneously polynomial time, planarity-preserving,
and parsimoniously reducible to each of the basic CNF satisfiability
problems listed in Table 1. 
( Previously, {\sc 3Sat} was only known to be 
so reducible to the problem {\sc 1-Ex3Sat} \cite{DF86}.)

\item
There exist polynomial time, weakly parsimonious, 
and planarity-preserving reductions from the problem 
{\sc 1-Ex3MonoSat}
to several graph  problems including
{\sc Minimum Vertex Cover, Minimum Dominating Set,}
and {\sc Minimum Feedback Vertex Set}. 

\item
Using the results 1,2 and 3, variants of known reductions, and
new reductions, we show that for all
of the problems {\sc Pl-$\Pi$} in Table 1,
the problem {\sc \#Pl-$\Pi$} is {\sc \#P}-hard.
Similarly, we show that for many of the problems in Table 1,
the problems {\sc Ambiguous Pl-$\Pi$} and {\sc Unique Pl-$\Pi$} 
are {\sf NP}-hard and {\sf D$^P$}-hard, respectively.

\item
All of the {\sf \#P}-hardness results for planar counting problems in Table 1
can easily be shown to hold, even when the input to the problem consists of 
\begin{romannum}
\item[{}]
 a formula $f$ and one of its satisfying assignments, a graph
$G$ and one of its minimum 
vertex covers, a graph $G$ and  one of its 
dominating sets, etc.
\end{romannum}
\end{remunerate}
Thus the {\sf \#P}-hardness of the problems {\sc \#Pl-$\Pi$} in
Table 1 is not simply a corollary of the {\sf NP}-hardness of the problem
{\sc Pl-$\Pi$}, since the  problems {\sc \#Pl-$\Pi$} are {\sf \#P}-hard, 
even when restricted
to sets of instances for which the problems {\sc Pl-$\Pi$} are trivially
polynomial time solvable. Moreover, quoting from \cite{VV85} 
we note that --
\begin{romannum}
\item[{}]
``Whether the number of solutions of all 
{\sf NP}-complete problems are nevertheless
polynomial time interreducible (i.e., whether 
{\sf NP}-completeness implies {\sf \#P}-completeness) is still open."
\end{romannum}

Corollaries of our results and their proofs include the following:
\begin{remunerate}
\item
The problems {\sc \#1-Valid 3Sat} and {\sc \#1-Valid Pl-3Sat} are 
{\sf \#P}-complete.
(It is trivially seen that every instance of the problem {\sc 1-Valid 3Sat}
is satisfiable by the assignment making all 
variables equal to 1. See Definition \ref{def2.9}. )

\item
The problems {\sc Ambiguous 1-Valid 3Sat} and 
{\sc Ambiguous 1-Valid Pl-3Sat} are {\sf NP}-complete.
The problems {\sc Unique 1-Valid 3Sat} and 
{\sc Unique 1-Valid Pl-3Sat} are {\sf Co-NP}-complete.

\item
Assuming {\sf P $\neq$ NP}, there are no $\epsilon$-approximation algorithms
for the problems of maximizing or minimizing 
a linear objective function subject
to a planar system of inequalities over the integers.
\end{remunerate}

\noindent
Table 1 gives a summary of our {\sf \#P}-hardness results.
The rest of the paper is organized as follows. Section \ref{sec:def} 
contains definitions and preliminaries. 
Section \ref{sec:3sat} discusses the complexity of {\sc \#Pl-3Sat}
and other basic CNF satisfiability problems.
These problems are  used to prove the
{\sf \#P}-hardness of other problems discussed in the subsequent sections.
Section \ref{sec:graph} discusses the
 complexity of various counting
problems for planar graphs.
Section \ref{sec:unique}
contains the ambiguous  and unique satisfiability results
and the result on the non-approximability of the objective
functions of  integer linear programs.
Finally, \S\ref{sec:conclusions} 
consists of conclusions and open problems.


\vspace{.5in}
\baselineskip = 0.8\normalbaselineskip

\begin{center}
\begin{tabular}[btp]{||c|c|c|c||}
\hline
S.No &   Problem       &  Decision Problem  &  Counting Version  \\
     &               &  ({\sf NP}-complete)  &  {\sf \#P}-hard    \\
\hline
1    &   {\sc Pl-3Sat}    &  \cite{DL82}        &   *            \\
2    &   {\sc Pl-Ex3Sat}    &  *        &   *            \\
3    &   {\sc Pl-1-3Sat}      &  \cite{DF86}        &   *                  \\
4    &   {\sc Pl-1-Ex3Sat}   &   \cite{DF86}     &   *                  \\
5    &   {\sc Pl-1-Ex3MonoSat} &  *                &   *               \\ 
\hline
6    &   {\sc Pl-Vertex Cover} &   \cite{GJS76}    &  *        \\
7    & {\sc Pl-Hamiltonian Circuit} & \cite{DL82}  &  \cite{SP86}    \\
8   &  {\sc Pl-Dominating Set} &  \cite{GJS76}     &  *      \\
9    & {\sc Pl-Feedback Vertex Set} & \cite{GJS76}  &  *  \\
10   &  {\sc Pl-3-Coloring} &  \cite{GJS76}  &  \cite{HMRS93}  \\
11   & {\sc Pl-Graph Homomorphism}  &  \cite{GJ79}  &  \cite{HMRS93}  \\
     &    and {\sc Onto Homomorphism} & & \\
12   & {\sc Pl-Subgraph Isomorphism} & \cite{GJ79} & *   \\
13  & {\sc Pl-Clique Cover}  &         \cite{DF86}        &   *     \\
14   &  {\sc Pl-Hitting Set}   & \cite{DF86}                 &    *  \\
15   &  {\sc Pl-X3C}         &  \cite{DF86}       &   *   \\
16   &  {\sc Pl-Partition Into Triangles}  &  \cite{DF86}  &  *  \\
17   &  {\sc Pl-Partition Into Claws}      &  \cite{DF86}  &  *    \\
\hline
\end{tabular}
\end{center}
\begin{center}
{\bf Table 1: Summary of {\sf NP}- and \#{\sf P}-hardness 
results for planar instances. The third 
column summarizes the decision complexity of the problems while the fourth 
column summarizes the complexity of the counting versions. 
A star (*) denotes  result obtained in this paper. The numbers in square
brackets is the reference where the corresponding result is proved.}
\end{center}

\section{Definitions and preliminaries}\label{sec:def}
In this section we  review  the basic definitions and notation
used in this paper. Additional definitions can be found in 
\cite{DF86,GJ79,Pa94,Sa80}.

\begin{definition}\label{def2.1}
A {\em search problem} $\Pi$ consists of a set  $D_{\Pi}$ of objects called
{\em instances}, and for each instance $I \in D_{\Pi}$, a set $S_{\Pi}[I]$
of objects  called {\em solutions} for $I$. An algorithm is said to 
{\em solve} a search problem $\Pi$ if, given $I \in D_{\Pi}$ as input, the 
algorithm outputs {\bf no} if $S_{\Pi}[I] = \phi $ and outputs an
$s \in S_{\Pi}[I]$ otherwise.  
\end{definition}

\begin{definition}\label{def2.2}
The {\em enumeration problem} associated with a search
problem $\Pi$ is the problem of determining, given $I \in D_{\Pi}$, the
cardinality of $S_{\Pi}[I]$.  
\end{definition}

\begin{definition}\label{def2.3}
The class {\sf \#P} consists of all enumeration problems
associated with search problems $\Pi$ such that
there is a non-deterministic algorithm for solving $\Pi$ such that, for all
$I \in D_{\Pi}$, the number of distinct 
accepting sequences for $I$ by the algorithm equals $|S_{\Pi}[I]|$
and the length of the longest accepting computation of the algorithm on 
$I \in D_{\Pi}$ is bounded by a 
polynomial in the length of $I$. 
\end{definition}

\begin{definition}\label{def2.4}
A reduction \cite{GJ79}
$f: D_{\Pi} \rightarrow D_{\Pi '}$ is {\it parsimonious} if and only if 
 $\forall I \in D_{\Pi}$ the number of solutions of $I$
is equal to the number of solutions of $f(I)$. 
\end{definition}

\begin{definition}\label{def2.5}
A reduction $f$ is
{\it weakly parsimonious} if and only if $|S(I)| = g(I) |S(f(I))|$, where
$|S(I)|$ and $|S(f(I))|$ denote the number of solutions of $I$ and $f(I)$ 
respectively and 
$g(I)$ is a  polynomial time computable function represented using binary
notation. 
\end{definition}

An enumeration problem is said to be {\sf \#P-hard} if each problem in 
{\sf \#P}
is polynomial time parsimoniously or weakly parsimoniously 
Turing reducible to it, If in addition, the enumeration
problem is in {\sf \#P}, 
the problem is said to be {\sf \#P-complete}.

\begin{definition}\label{def2.6}
{\sc \#3Sat} 
is the problem of computing the number of satisfying assignments of
a Boolean formula $F$ in conjunctive normal form with at most three literals
per clause. 
\end{definition}

The following basic results on the 
complexity of counting problems are used in this paper.

\begin{theorem}\cite{GJ79,Va79a}
The problems  {\sc \#3Sat}  and {\sc \#Graph 3-Coloring} 
are  {\sf \#P}-complete. 
\end{theorem}

\begin{definition}\label{def2.7}
The {\em bipartite graph} associated with a CNF satisfiability problem
is defined as
follows: The clauses and  variables in a formula are in one to one
correspondence with the vertices of the graph. There is an edge between a 
clause node and a variable node if and only if the  variable appears
in  the clause.
A CNF formula is {\it planar} if and only if 
its associated bipartite graph is planar.
\end{definition}

\begin{definition}\label{def2.8}
\noindent
(1){\sc Ex3Sat} is the restriction of  the problem
{\sc 3Sat} to formulas in which each clause has exactly three literals.

\noindent
(2){\sc 1-3Sat} is the problem of determining if a CNF formula in which 
each clause has no more than 3 literals has a satisfying assignment
such that exactly one literal per clause is satisfied.

\noindent
(3){\sc 1-Ex3Sat} is the problem of determining if a CNF formula in which 
each clause has exactly 3 literals has a satisfying assignment
such that exactly one literal per clause is satisfied.

\noindent
(4){\sc 1-Ex3MonoSat} is the restriction of {\sc 1-Ex3SAT} to formulas having
no negated literals. 
\end{definition}

\noindent
{\bf Definition 2.9:} \cite{Sc78} 
\begin{definition}\label{def2.9} \cite{Sc78} 
A relation $R(x_1, x_2, \ldots, x_n)$ is {\em 1-valid}
 if and only if  $(1,1, \ldots , 1) \in R$.
A CNF formula $f$ is {\em 1-valid} if the formula is satisfied when
all the variables in the formula are set to true.
\end{definition}

\begin{definition}\label{def2.10}
Given  a Boolean formula $F$
and an assignment {\bf v} to the variables of $F$, the notation
{\bf v}$[F]$  denotes the value of $F$ under {\bf v}.
\end{definition}

\begin{definition}\label{def2.11}

\noindent
(1) {\sc Exact Cover By 3-Sets (X3C:)} An instance of this problem
consists of a set $X$ with $3m$ elements
and a collection $C$ of 3-element subsets of $X$. The question  is:
Does there exist a sub-collection 
$C'$ of $C$ such that  every element
of $X$ occurs in exactly one set in $C'$? 

\noindent
(2) {\sc Hitting Set:}
An instance of this problem consists of
a collection $C$ of subsets of a finite set $S$ and a positive
integer $K \leq |C|$. The question is: Is there a subset $S' \subseteq S$
with $|S'| \leq K$ such that $S'$ contains at least one element from
each subset in $C$? 
\end{definition}

As in the case of {\sc 3Sat}, we can associate
a bipartite graph $G=(S,T,E)$ with an instance of each of the above problems.
For example  {\sc Pl-X3C} is defined as follows:
Each element in $C$ has a corresponding vertex in $S$, each element in
$X$ has a corresponding vertex in $T$, and a vertex in $S$ is joined to
a vertex in $T$ if and only if the set corresponding to the vertex in $S$
contains the element corresponding to the vertex in $T$.

\begin{definition}\label{def2.12}

\noindent
(1) {\sc Dominating Set:}
An instance of this problem consists of an undirected graph $G=(V,E)$,
and an integer $K \leq |V|$. 
The question is: 
Is there a  dominating set of size at most $K$  in $G$,
i.e. is there a subset $V'$ of $V$, $|V'| \leq K$,  
such that for each $u \in V- V'$ 
there is a $v \in V'$ such that $(u,v) \in E$ ?

\noindent
(2) {\sc Clique Cover:}
An instance of this problem consists of an undirected graph $G=(V,E)$,
and an integer $K \leq |V|$. The question is: Is there a clique cover
of size at most $K$  in $G$, i.e. can 
the graph be partitioned into at most $K$ sets of nodes such that each set is
a clique ? 

\noindent
(3) {\sc Partition into Claws:}
An instance of this problem consists of an undirected graph 
$G=(V,E)$, $|E| = m$. 
The question is: 
Is there a way to partition the edges of the graph
into sets $E_1,\ldots, E_s$, $s = m/3$, 
such that each $E_i$ induces a subgraph isomorphic to $K_{1,3}$
(i.e. a claw)? 

\noindent
(4) {\sc Feedback Vertex Set:}
An instance of this problem consists of a directed graph $G = (V,E)$ and an
integer $K \leq |V|$. The
question is: Is there a feedback vertex set of size at most $K$  in $G$,
i.e. does there exist a subset $V'$ of  $V$ of size at most
$K$ such that $V'$ contains at least one vertex from every cycle in $G$ ?
\end{definition}


\begin{definition}\label{def2.13}
Let $\Pi$ denote a CNF satisfiability problem. Then the associated
{\em ambiguous version} of $\Pi$, denoted by {\sc Ambiguous-$\Pi$},
is the problem of determining, given an instance $I$ of $\Pi$ and
an assignment {\bf v} to the variables of $I$ such that {\bf v}$[I]= 1$,
if there is an additional assignment of values to the variables of $I$
satisfying $I$.  The associated
{\em unique version} of $\Pi$, denoted by
{\sc Unique-$\Pi$}, is the problem
of determining if the given instance $I$ of $\Pi$ has  exactly
one satisfying assignment. 
\end{definition}

More generally, let $\Pi$ be any decision problem. Henceforth when
applicable, we denote the restriction of $\Pi$ to planar instances
by {\sc Pl-$\Pi$}, the
counting version of $\Pi$ by \#$\Pi$,
the ambiguous version of $\Pi$ by {\sc Ambiguous-$\Pi$}, and 
the unique version of the problem by {\sc Unique-$\Pi$}. For example, 
recalling Definition \ref{def2.11} and the discussion immediately following it,

\noindent
(i){\sc Pl-X3C} is the problem {\sc X3C} 
restricted to instances $(X,C)$ for which
the bipartite graph  $G=(S,T,E)$ is planar.

\noindent
(ii){\sc \#X3C} is the problem of computing, 
given $(X,C)$, the number of distinct
subsets $C' \subseteq C$ such that each element of $X$ occurs in exactly
one set in $C'$.

\noindent
(iii) {\sc Ambiguous-X3C} is the problem of determining, given $(X,C,C')$ where
$C' \subseteq C$ and each element of $X$ occurs in exactly one set in $C'$,
if there exists another subset $C'' \subseteq C$ 
that is an exact cover of the elements in $X$.

Finally henceforth by {\it reduction}, we mean a  polynomial time many one 
reduction.

\section{ Basic planar counting problems}\label{sec:3sat}
In this section, we prove that the problem
{\sc Pl-3Sat} has a parsimonious crossover box. Specifically, we show that
the crossover box in \cite{DL82} is parsimonious.
One immediate corollary is that the problem {\sc \#Pl-3Sat}
is {\sf \#P}-hard. 
We also prove that the problem
{\sc 3Sat} is  planarity preserving and parsimoniously
reducible to each of the basic {\sc Sat} problems listed in Table 1.

\begin{definition}\label{def3.1}
A {\bf crossover box} for a satisfiability problem $\Pi$ is a formula $f_c$
with four distinguished variables $a$, $a_1$, $b$, $b_1$, which
can be laid out on the plane with the distinguished variables on the
outer face, such that
\begin{remunerate}
\item the {\bf old variables} $a$ and $b$ are opposite to the corresponding 
{\bf new variables} $a_1$ and $b_1$,
\item each assignment to $a$ and $b$ can be extended to a satisfying assignment
of $f_c$,
\item for any satisfying assignment of 
$f_c$, $a \equiv a_1$ and $b \equiv b_1$.
\end{remunerate}
The crossover box is {\bf parsimonious} if and only if for each assignment to
the old variables, there is a exactly one extension of this assignment to the 
variables in the crossover box such that $f_c$ is satisfied.
\end{definition}

\begin{theorem}\label{cross}  
The problem {\sc 3Sat} has a parsimonious planar crossover box.
Hence, {\sc 3Sat} is parsimoniously reducible to {\sc Pl-3SAT} and 
{\sc \#Pl-3Sat} is {\sc \#P}-complete.
\end{theorem}
\begin{proof}
The crossover box described here is the same as the one given in
\cite{DL82}. Here we prove that the crossover box also preserves the number
of solutions.
For expository purposes, we describe the crossover box in two steps.
The first step is to consider the following formula:\\
\[f_{c} = (\overline{a_2} + \overline{b_2} + \alpha) \wedge 
(a_2 + \overline{\alpha}) \wedge (b_2 + \overline{\alpha}) \bigwedge \]
\[ (\overline{a_2} + b_1 + \beta) \wedge 
(a_2 + \overline{\beta}) \wedge (\overline{b_1} + 
\overline{\beta}) \bigwedge\]
\[(a_1 + b_1 + \gamma) \wedge (\overline{a_1} + \overline{\gamma}) 
\wedge  (\overline{b_1} + \overline{\gamma}) \bigwedge \]
\[ (a_1 + \overline{b_2} + \delta) \wedge 
(\overline{a_1} + \overline{\delta}) \wedge 
(b_2 + \overline{\delta}) \bigwedge \]
\[(\alpha + \beta + \gamma + \delta ) \bigwedge \]
\[(\overline{\alpha} + \overline{\beta}) \wedge
(\overline{\beta} + \overline{\gamma}) \wedge
(\overline{\gamma} + \overline{\delta}) \wedge
(\overline{\delta} + \overline{\alpha}) \bigwedge \]
\[(a_2 + \overline{a}) \wedge (a + \overline{a_2}) \wedge
(b_2 + \overline{b}) \wedge (b + \overline{b_2})\]

Following \cite{DL82}, we give a intuitive explanation of formula $f_c$.
Clauses 1, 2 and 3 imply that $(\alpha \leftrightarrow (a_2 \wedge b_2))$,
clauses 4, 5 and 6 imply $(\beta \leftrightarrow (a_2 \wedge \overline{b_1}))$.
clauses 7, 8 and 9 imply
$(\gamma \leftrightarrow (\overline{a_1} \wedge \overline{b_1})),$
clauses 10, 11 and 12 imply
$(\delta \leftrightarrow (\overline{a_1} \wedge b_2))$.
Clause 13 (the four literal clause) implies that at least one of 
$\alpha$, $\beta$, $\gamma$ or $\delta$ is true.
Clauses 14, 15, 16, 17 imply that 
$(\alpha + \gamma) \rightarrow ( \overline{\beta} \wedge \overline{\delta})$ 
and
$( \beta + \delta)  \rightarrow (\overline{\alpha} \wedge \overline{\gamma})$.
Finally, clauses 18 and 19 imply $(a \leftrightarrow a_2),$ and
$(b \leftrightarrow b_2)$.
It can now be verified that the formula $f_c$ implies 
$(a_1 \leftrightarrow a)$ and $(b_1 \leftrightarrow b)$.
For example, consider an assignment {\bf v} such that
{\bf v}[$a_1$] $ = $ {\bf v}[$b_1$] $ = 0$. Then $f_c$ implies that
{\bf v}[$\beta$] $ = $ {\bf v}[$\delta$] $ = $ {\bf v}[$\alpha$] $ = $
{\bf v}[$a_2$] $ = $ {\bf v}[$b_2$] $ = 0$ and {\bf v}[$\gamma$] $ = 1$.  
We leave it to the reader to verify the other 3 cases.
Thus, in any satisfying assignment to $f_c$, the new variables
$a_1, a_2, b_1, b_2, \alpha, \beta, \gamma$ and $\delta$ 
are {\em functionally} dependent on $a$ and $b$. In other words,
given an assignment to the variable $a$ and $b$ there is a
unique way to extend this assignment so as to satisfy all the clauses
in $f_c$. Thus $f_c$ is a parsimonious crossover box. Even though, 
the formula itself is a parsimonious planar crossover box, it is unsuitable for
a reduction to {\sc Pl-3Sat} because it has one clause with 
four literals, namely $(\alpha + \beta + \gamma + \delta )$. 
The second step is to 
obtain the formula $f_{c}'$ by replacing
this clause with the formula
$(\alpha + \delta + \xi) \wedge (\overline{\xi} + \beta + \gamma)$.
The planarity of the formula $f_{c}'$ is demonstrated in 
Figure~\ref{crossoverbox1.fig}.
This step also preserves numbers of satisfying assignments as demonstrated
by the following claim.

\begin{claim}
(1) Exactly one of $\alpha, \beta, \gamma, \delta$ is true in any satisfying 
assignment to $f_c$. 
(2) $\xi$ is functionally dependent on $\alpha, \beta, \gamma$ and  $\delta$.
Thus a satisfying assignment for $f_c$ can be extended in a unique way
to a satisfying assignment to the formula $f_{c}'$. 
\end{claim}

\begin{proof}
We prove the claim for the case when $\alpha$ is true. 
The other three cases are similar.
As already discussed, clauses 14, 15, 16, 17 imply that 
$(\alpha + \gamma) \rightarrow ( \overline{\beta} \wedge \overline{\delta})$ 
and
$( \beta + \delta)  \rightarrow (\overline{\alpha} \wedge \overline{\gamma})$.
Consider a satisfying assignment {\bf v} such that
{\bf v}[$\alpha$] $ = 1$. Then the above discussion implies that
{\bf v}[$\beta$] $ = $ {\bf v}[$\delta$] $ = 0$.
Now, since $(\beta \leftrightarrow (a_2 \wedge \overline{b_1}))$ and
{\bf v}[$\beta$] $ = 0$, we have
{\bf v}[$a_2$] $ = 1 $ and  {\bf v}[$b_1$] $ = 1$. Since
$(\gamma \leftrightarrow (\overline{a_1} \wedge
\overline{b_1}))$ and {\bf v}[$b_1$] $ = 1$ it implies that
{\bf v}[$\gamma$] $ = 0$. This forces {\bf v}[$\xi$] $= 0$. 
\qquad \end{proof}

Thus, the satisfying assignments to  $f_{c}'$ 
satisfy $f_{c}$. It can now be seen that
in a satisfying assignment to the variables of $f_c'$, 
the values of $a_1$, $a_2$, $b_1$, $b_2$, $\alpha$, 
$\beta$, $\gamma$, $\delta$, and $\xi$ are all functionally dependent on
$a$ and $b$. Therefore $f_{c}'$ is a parsimonious crossover box.

\begin{figure}[tbp]
\centerline{\epsffile{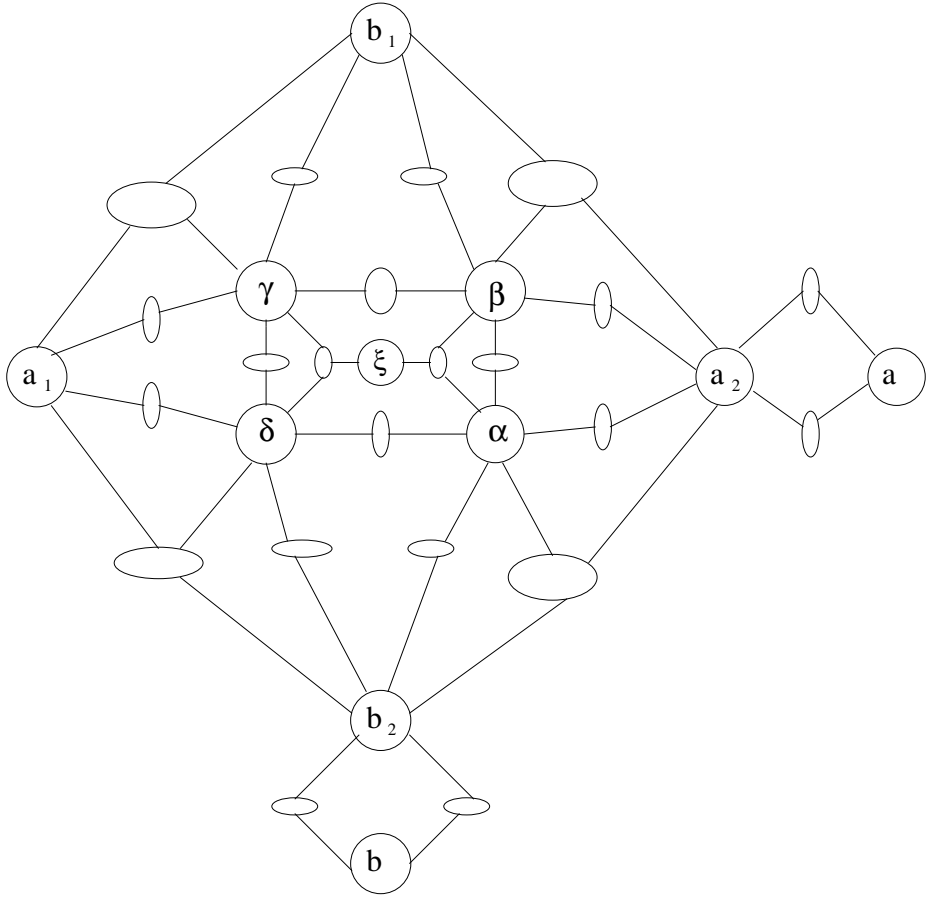}}
\caption{The parsimonious planar crossover box for {\sc 3Sat}. The clauses are
denoted by ellipses and the variables are denoted by circles.}
\label{crossoverbox1.fig}
\end{figure}

\begin{figure}[tbp]
\centerline{\epsffile{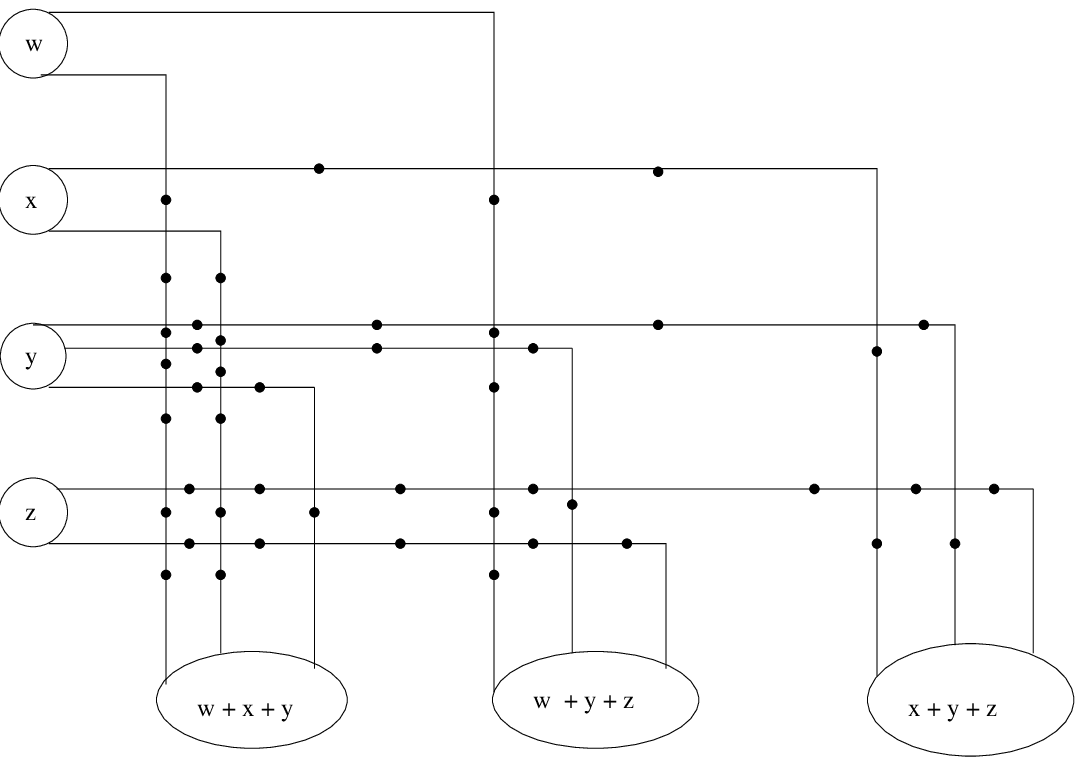}}
\caption{Figure showing how new variables are introduced 
in the bipartite graph for {\sc 3Sat} to obtain an instance of {\sc Pl-3Sat}.
In the example, the instance of {\sc 3Sat} is given by 
$F = (w + x + y) \wedge (w + y + z) \wedge  (x + y + z)$.
The circles denote original variables, ellipses denote the original
clauses and the black dots denote new variables added. Each 
crossover of edges is replaced by a crossover box  shown in 
Figure \ref{crossoverbox1.fig}.}
\label{fig31.fig}
\end{figure}

We can now describe the reduction given in \cite{DL82} from {\sc 3Sat} to 
{\sc Pl-3Sat}. For any 3CNF formula, lay the formula in the plane,
possibly with certain edge pairs crossing over at ``crossover points".
This layout is a planar graph with vertex set consisting of the
variables, clauses and the  crossover points.
In this layout, we add a new variable node on the edge between two crossover
points or between a crossover point and a clause node as shown in 
Figure~\ref{fig31.fig}.
The resulting graph is bipartite where the first set of nodes consists
of the variable nodes and the second set consists of the clause nodes and the
crossover points. Each edge is between a variable and a crossover point, or
between a variable and a clause. Also, each crossover point has
four distinct variables as neighbors. We now replace each crossover point
with the crossover box in Figure~\ref{crossoverbox1.fig} 
where $a_2$, $b_2$, $\alpha$, $\beta$, $\gamma$, $\delta$, and $\xi$ are
given distinct names in each replacement. Here $a$, $b$, $a_1$ and $b_1$
are identified with the neighbors of the crossover point in cyclic order in
the layout.
The new layout is planar and
the new formula has the same number of solutions as the original, since
in any satisfying assignment the new variables are functionally dependent on 
the old. \qquad \end{proof}

Next, we strengthen   Theorem \ref{cross} by 
showing that  counting the number of satisfying assignments of a planar
3CNF formula is {\sf \#P}-complete, even when the input consists of a planar
3CNF formula $F$ and an assignment {\bf v}$[F]$ to the variables of $F$ which 
satisfies $F$. 

\begin{theorem}\label{ambi}
Given an instance $F$ of {\sc Pl-3Sat} and an assignment 
{\bf v} to the variables
of $F$ such that {\bf v}$[F] = 1$, the problem of
counting the number of satisfying assignments of $F$  is {\sf \#P}-complete.
\end{theorem}
\begin{proof}

\noindent
{\bf Step 1:}
Given an arbitrary planar 3CNF formula $f(x_1, x_2, \ldots, x_n)$, 
we first construct a new formula\\
$f_1(x_1, x_2, \ldots, x_n, x_{n+1})$  where 
$f_1 = (f \wedge x_{n+1}) \bigvee 
[\overline{x_{1}} \wedge
\overline{x_2} \wedge \ldots \overline{x_n}  \wedge \overline{x_{n+1}}]$.

Obviously, an assignment {\bf v},  such that,
{\bf v}[$x_1$] $ \ = \ $ {\bf v}[$x_2$]  $ \ = \ \cdots  = $ 
{\bf v}[$x_n$] $ \ = \ $ {\bf v}[$x_{n+1}$] $ \ = \ 0$  
satisfies $f_1$. Hence $f_1$ is always satisfiable.
Also, the number of satisfying 
assignments of $f_1$ is one more than the number of satisfying assignments
of $f$.
Therefore, knowing the the number of satisfying
assignments of $f_1$ tells us the number of satisfying assignments
of $f$.

\noindent
{\bf Step 2:}
Convert $f_1(x_1, x_2, \ldots, x_n, x_{n+1})$, into an equivalent  3CNF
formula \\
$f_2(x_1, x_2, \ldots, x_n, x_{n+1}, x_{n+2}, \ldots, x_{n+p})$,
where $x_{n+2}, \ldots, x_{n+p}$ are new variables.
This is done in the standard way as follows:
Obtain a parse tree of $f_1$. For each non-leaf node in the 
parse tree introduce new variables $y_1, y_2, \ldots, y_m$, where
$y_m$ is the variable corresponding to the root of the parse tree.
Each node of the parse tree corresponds to an operator applied to
one/two inputs. Let the children of a non leaf node $i$ be $j$ and $k$.

\noindent
{\bf Case 1:} If the operator at node $i$ is an {\sf AND} operator, 
construct a new 3CNF formula equivalent to the formula
$ y_i \equiv (y_j \wedge y_k)$.

\noindent
{\bf Case 2:} If the operator is an {\sf OR} operator, 
construct a 3CNF formula equivalent to the formula
$ y_i \equiv (y_j + y_k)$.

\noindent
{\bf Case 3:} If the operator is an {\sf NOT} 
operator then construct a 3CNF formula equivalent to the formula
$ y_i \equiv \overline{y_j}$.

The final 3CNF formula $f_2$ is a 
conjunction of all the 3CNF formulas along with
$y_m$. Now it is easy to see that $f_2$ is satisfiable if and only if $f_1$ is
satisfiable and the reduction is parsimonious. (The new variables are 
functionally dependent on the old.)

\noindent
{\bf Step 3:}
Next, lay out $f_2$ on a plane 
and replace each crossover in the layout by  the
crossover box described in Theorem~\ref{cross}. Let $f_3$ be the resulting
planar 3CNF formula.
By Theorem~\ref{cross}, the reduction from $f_2$ to $f_3$ is parsimonious.
Thus the number of satisfying assignments of $f_3$ is 1 more than the 
number of satisfying assignments of $f$. The theorem now follows.
\qquad \end{proof}

Next we extend this result to prove that counting the number of
satisfying assignments of a {\em 1-valid} planar 3CNF formula is 
{\sf \#P}-hard.

\begin{theorem}\label{1-valid}
The problem {\sc \#1-Valid Pl-3Sat} is {\sf \#P}-complete.
\end{theorem}
\begin{proof}
Given an arbitrary {\sc Pl-3CNF} formula  $f(x_1, x_2, \ldots, x_n)$
and a satisfying assignment {\bf v} such that {\bf v}[f] $ = 1$,
we construct a new formula $f_1(x_1, x_2, \ldots, x_n, y_1, \ldots y_p)$
where $y_1, \ldots, y_p$ are new variables.
The formula $f_1$ is constructed as follows: Let
$x_{l_1}, x_{l_2}, \ldots x_{l_p}$ be the variables in $f$ such that
{\bf v}[$x_{l_1}$] $ = $ {\bf v}[$x_{l_2}$] $ =  \cdots \  = $ 
{\bf v}[$x_{l_p}$] $ = 0$.
Then
replace  $\overline{x_{l_i}}$ by $y_i$ and  $x_{l_i}$ by $\overline{y_i}$,
$1 \leq i \leq p $. 
Obviously, the formula $f_1$ is 1-valid and the reduction is parsimonious.
\qquad \end{proof}

Karp and Luby \cite{KL83} presented randomized fully polynomial time 
approximation schemes for several {\sf \#P}-complete problems, including
{\sc \#DNF}. Since then, substantial research has been done
in the area of approximation algorithms for various counting problems. 
Saluja et al. \cite{SST92} give a logical characterization of the counting
problems that have a polynomial time random approximation scheme.
Our {\sf \#P}-hardness of {\sc \#Pl-3Sat} and other problems immediately
raise the question of approximating the optimal values of these
counting problems.  
The parsimonious reduction from {\sc 3Sat} to {\sc Pl-3Sat},
implies that,
given a deterministic polynomial time algorithm $A$ to
approximately count the number of satisfying assignments of a planar 3CNF
formula, we can construct a deterministic polynomial time
algorithm $A'$  with the same performance guarantee,
to  approximately count the number of satisfying assignments
of an arbitrary 3CNF formula.

Intuitively Theorem~\ref{cross} and the above observation mean that
counting the number of satisfying assignments of a planar 3CNF formula is 
{\em as hard as}
counting the number of satisfying assignments of an arbitrary 3CNF formula
with respect to polynomial time reducibility. We remark that the
result  holds even for {\sc 1-Valid Pl-3CNF} formulas.

Next, we prove the {\sf \#P}-hardness of other basic satisfiability problems.
First we prove two lemmas:

\begin{lemma}\label{ex-one}
Let $F$ be the planar monotone formula
$(c+d+e) \wedge (c+e+f) \wedge (d+e+f)$.
Then there is a unique satisfying assignment {\bf v} to the variables of $F$
such that each clause has exactly one true literal,
namely, 
{\bf v}$[c] \ = \ $  {\bf v}$[d]  \ = \ $ {\bf v}$[f] \ =  0 \ $;
 and {\bf v}$[e]  \ = 1 \ $.

\end{lemma}
\begin{proof} By inspection. \qquad \end{proof}

\begin{lemma}\label{unique}
The following {\sc Ex3CNF} formula is planar and has exactly one satisfying 
assignment,
namely the assignment {\bf v} defined by 
{\bf v}$[x_i] = 0 (1 \leq i \leq 9)$:\\
$G(x_1, \ldots , x_9 ) = 
(\overline{x}_{1} + \overline{x}_{2} + \overline{x}_{3} )\wedge 
(\overline{x}_{1} + x_2 + x_7 )\wedge 
(\overline{x}_{2} + x_3 + x_8 )\wedge 
(\overline{x}_{3} + x_1 + x_9 ) \wedge 
(\overline{x}_{4} + x_1 + x_7 )\wedge
(\overline{x}_{5} + x_2 + x_8 ) \wedge
(\overline{x}_{6} + x_3 + x_9 ) \wedge 
(\overline{x}_{7} + x_5 + x_8 )\wedge
(\overline{x}_{8} + x_6 + x_9 ) \wedge
(\overline{x}_{9} + x_4 + x_7 ) \wedge
(\overline{x}_{1} + \overline{x}_{4} + \overline{x}_{9} ) \wedge
(\overline{x}_{2} + \overline{x}_{5} + \overline{x}_{7} ) \wedge
(\overline{x}_{3} + \overline{x}_{6} + \overline{x}_{8} ) \wedge 
(\overline{x}_{7} + \overline{x}_{8} + \overline{x}_{9} )$.
\end{lemma}

\begin{figure}[tbp]
\def\epsfsize#1#2{0.85#1}
\centerline{\epsffile{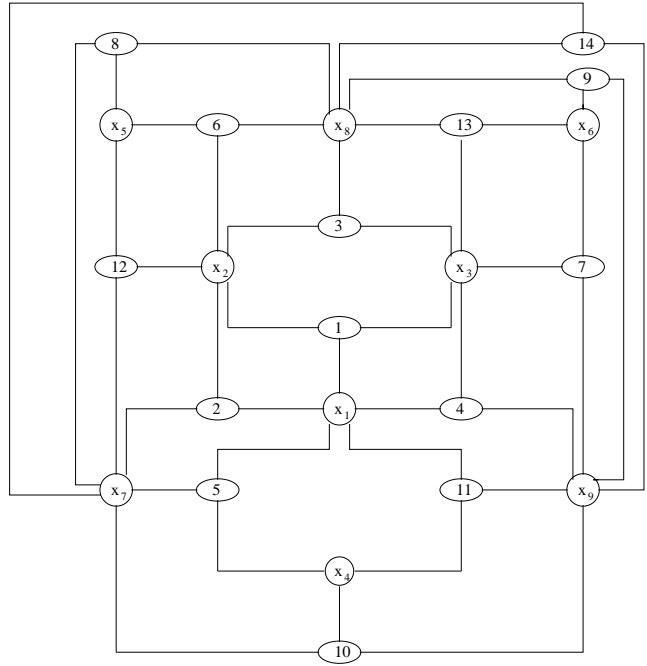}}
\caption{{\sc Pl-Ex3Sat} formula as described in Lemma \ref{unique}.
Ellipses denote the clauses in $G$. 
The clauses are numbered in the order in which they
appear in Lemma \ref{unique}.}
\label{ex3Sat.fig}
\end{figure}

\begin{proof} Planarity is demonstrated by Figure \ref{ex3Sat.fig}.
The given assignment {\bf v} satisfies $G$ since each clause of $G$
contains a negated literal.

Suppose $G$ is True. Let {\bf u} be any truth assignment to the variables 
in $G$, Then, 
$x_7 \rightarrow x_8 $ by clauses 6, 8, and 12; 
$x_8 \rightarrow x_9 $ by clauses 7, 9, and 13; 
$x_9 \rightarrow x_7 $ by clauses 5, 10, and 11.  
Therefore, 
$(x_7 + x_8 + x_9 ) \rightarrow (x_7 \wedge x_8 \wedge x_9 )$.   
Hence by clause 14, 
{\bf u}[$x_7$] $\ = \ $  {\bf u}[$x_8$]  $\ = \ $ {\bf u}[$x_9$] $= 0$.
Given this 
$x_1 \rightarrow x_2 $ by clause 2; 
$x_2 \rightarrow x_3 $  by clause 3; and 
$x_3 \rightarrow x_1 $ by clause 4.  
Therefore, 
$(x_1 +  x_2 +  x_3 ) \rightarrow (x_1 \wedge x_2 \wedge  x_3 ).$  
Hence by clause 1, 
{\bf u}[$x_1$] $\ =\ $ {\bf u}[$x_2$] $\ =\ $ {\bf u}[$x_3$] $\ =\  0.$  
But this implies that 
{\bf u}[$x_4$] $\ =\ $ {\bf u}[$x_5$] $\ =\ $ {\bf u}[$x_6$] $\ =\ 0 $ 
by clauses 5, 6, and 7.  Hence in any satisfying assignment  {\bf u} of
$G$, {\bf u}[$x_i$] $\ =\ 0 \ (1  \leq  i  \leq   9).$  
\qquad \end{proof}

Next we give planarity-preserving parsimonious reductions from {\sc 3Sat} to
the basic {\sc Sat} problems listed in Table 1. Without loss of
generality, 
we assume that the given instance of the CNF formula does not have any single
literal clause. 
\begin{theorem}\label{parsi}
There exist planarity-preserving parsimonious reductions from {\sc 3Sat} to
each of the following problems:
{\sc Ex3Sat, 1-3Sat, 1-Ex3Sat, 1-Ex3MonoSat}
and {\sc X3C}.
\end{theorem}
\begin{proof}
{\sc 3Sat} $\rightarrow$ {\sc Ex3Sat}:

Let $f$ be a 3CNF formula with clauses 
$c_j (1 \leq j \leq k).$  For $1  \leq j \leq  k,$ let 
$c_j'$ be $c_j $, if $ c_j $ is a three-literal clause. 
If $c_j = (l_{j1} +  l_{j2})$,
let $c_j'$ be 
$(l_{j1} + l_{j2} + x_9^j ) \wedge G(x_1^j, \ldots ,x_9^j ). $
\par\noindent

$G$ is defined 
as in Lemma \ref{unique} 
and $x_i^j (1 \leq i \leq 9)$ are distinct new variables. 
Let $\displaystyle{g = \bigwedge_{j=1}^k  c_j'}$.
Then by Lemma \ref{unique},
Figure \ref{ex3Sat.fig}
and direct inspection of the definitions of the formulas 
$c_j' $, the reduction mapping $f$ into $g$ is seen to be  
a planarity-preserving and parsimonious
reduction of the problem {\sc 3Sat} to the 
problem {\sc Ex3Sat}.  

\noindent
{\sc 3Sat} $\rightarrow$ {\sc 1-Ex3Sat}\footnote{Although they claim
to have a parsimonious reduction, the reduction actually given in
\cite{DF86} is {\bf not} parsimonious.}:
Let $f$ be a 3CNF formula 
with clauses $c_j(1\leq j\leq m)$.\\
(1)
For each three-literal clause $c_j = (z_p + z_q + z_r)$ of $f$, let
$c_j' = (z_p + u^j + v^j) \wedge (\overline{z_q} + u^j + w^j) \wedge $
$ (v^j + w^j + t^j) \wedge (\overline{z_r} + v^j + x^j), $
where $u^j, \; v^j, \; w^j, \; t^j$ and $x^j$ are distinct new variables
local to $c_j'$.\\
(2) For each two literal clause $c_j = (z_p + z_q)$ of
$f$, let 
$c_j' = (z_p + u^j + v^j) \wedge (\overline{z_q} + u^j + w^j) \wedge$
$ (v^j + w^j + t^j) \wedge (\overline{a^j} + v^j + x^j) \wedge $
$ (a^j + d^j + e^j) \wedge (a^j + e^j + f^j) \wedge (d^j + e^j + f^j),$\\
where $u^j,$ $v^j,$  $w^j,$ $t^j,$ $x^j,$  $a^j,$ $d^j$  and $e^j,$
are all new  variables local to $c_k'$. 	
Let $\displaystyle{f' = \bigwedge_{j=1}^m c_j'}$.

\begin{figure}[tbp]
\centerline{\epsffile{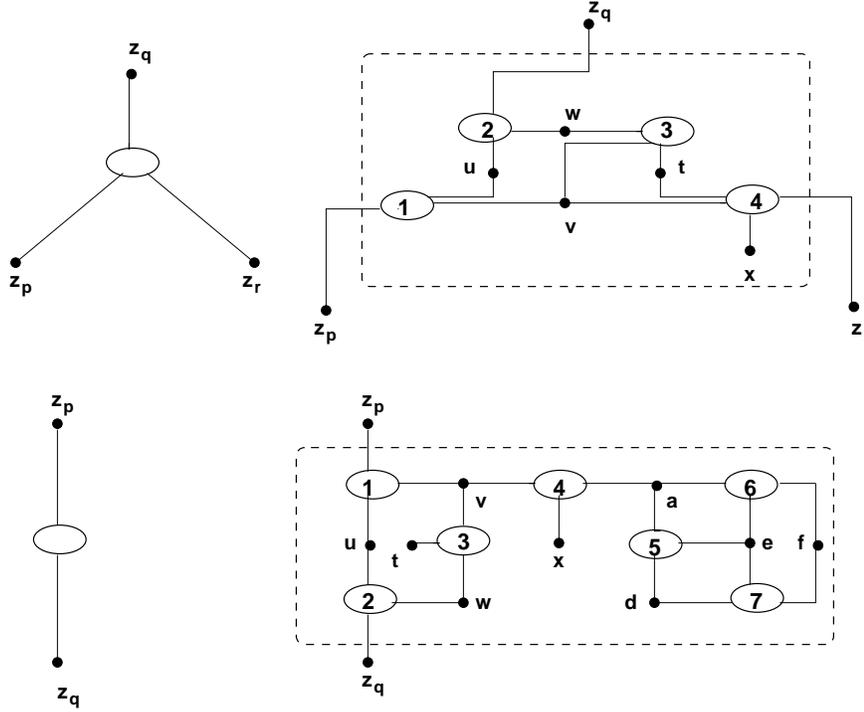}}
\caption{Figure illustrating the reduction from {\sc 3Sat} to {\sc 1-Ex3Sat}.
Figure \ref{fig3.fig}(a) shows how to transform a three literal clause.
Figure \ref{fig3.fig}(b) shows how to transform a two literal clause.
The clauses are numbered in the order in which they
appear in reduction outlined in proof of Theorem \ref{parsi}.
Note that the reduction is local replacement type reduction and hence
preserves planarity.}
\label{fig3.fig}
\end{figure}

To see the planarity of this reduction see Figure \ref{fig3.fig}.
We  claim that
$f'$ is exactly-one-satisfiable {\it if and only if}  
the original formula $f$
is satisfiable. Moreover the reduction is planarity-preserving and
parsimonious. To prove that the reduction is parsimonious, it suffices
to show the following two claims:
\begin{claim} 
No assignment of truth values to the variables of clause
$c_j\; (1 \leq j \leq m)$ of $f$ which does not satisfy $c_j$ 
can be extended to an assignment
of truth values to the variables of the formula $c_j'$ that
exactly-one satisfies $c_j'$. 
\end{claim}

\begin{proof}
Let $c_j \ =\ (z_p +  z_q +  z_r )$ 
and {\bf u} be an assignment to the variables such that 
{\bf u}[$z_p$] $ \ = \ $   {\bf u}[$z_q$] $ \ = \ $  
{\bf u}[$z_r$]  $ \ = \  0$.  Let {\bf v} be an exactly-one 
satisfying assignment to the variables of
$c_j'$ such that 
{\bf v}[$z_p$]  $ \ = \ $  {\bf v}[$z_q$] $ \ = \ $  {\bf v}[$z_r$] $ \ = \ 0$.
(i.e. {\bf v} is an extension of {\bf u}.)
Then, the clauses  $(\overline{z_q} +  u^j +  w^j )$ and 
$(\overline{z_r} +  v^j +  x^j )$ imply that 
{\bf v}[$u^j$] $ \ =\ $  {\bf v}[$w^j$] $ \ =\ $  {\bf v}[$v^j$] 
$ \  = \ $  {\bf v}[$x^j$] $ \ =\ 0.$  
It follows that {\bf v} does not exactly-one satisfy the clause 
$(z_p +  u^j +  w^j )$ of $c_j'$.   

Let $c_j \ =\ (z_p +  z_q )$  
and {\bf u} be an assignment to the variables such that 
{\bf u}[$z_p$] $ \ = \ $  {\bf u}[$z_q$]  $ \ = \ 0.$  
Let {\bf v} be an exactly-one 
satisfying assignment to the variables of
$c_j'$ such that {\bf v}[$z_p$] $ \ = \ $   {\bf v}[$z_q$] $ \ = \ 0.$ 
Then, the clauses $(z_p +  u^j +  v^j )$ and 
$(\overline{z_q} +  u^j +  w^j )$ imply that 
{\bf v}[$u^j$] $ \ = \ $  {\bf v}[$w^j$] $ \ = \ 0 $ and 
{\bf v }[$v^j$] $ \ = \ 1.$  But given this, the 
clause $(\overline{a^j} +  v^j +  x^j )$ 
implies that {\bf v}[$a^j$] $ \ = \ 1 $. 
Lemma \ref{ex-one} now implies that 
{\bf v} does not exactly-one satisfy $ c_j'$.
\qquad \end{proof}

\begin{claim}
For each satisfying assignment to the variables of the clause
$c_j, \; (1 \leq j \leq m)$ of $f$, there is exactly
one way the assignment can be extended to the variables of the formula 
$c_j'$ so as to exactly-one satisfy $c_j'$. 
\end{claim}
\begin{proof}
When
$ c_j \ =\ (z_p +  z_q +  z_r )$,
we need to verify  that the only exactly-one satisfying 
assignments $c_j' $ are the following:
\begin{remunerate}
\item  
$z_p \ =\  1,\ z_q  \ =\  0,\ z_r \ =\  0, \ u^j \ =\ 0, v^j \ =\  0,\ 
w^j \ =\ 0,\  t^j \ =\ 1, \ x^j \ =\ 0;$
\item
$z_p \ =\ 0,\ z_q \ =\ 1,\ z_r \ =\ 0,\  u^j \ =\ 1,\ v^j \ =\ 0,
 \ w^j \ =\ 0,\   t^j \ =\ 1, \ x^j \ =\ 0;$
\item
$z_p \ =\ 0,\ z_q \ =\ 0,\ z_r \ =\ 1,\  u^j \ =\ 0,\  v^j \ =\ 1,\  
w^j \ =\ 0,\ t^j \ =\ 0,\  x^j \ =\ 0;$
\item
$z_p \ =\ 1,\ z_q \ =\ 1,\ z_r \ =\ 0,\  u^j \ =\ 0,\ v^j \ =\ 0,\ 
w^j \ =\ 1,\  t^j \ =\ 0,\  x^j \ =\ 0;$
\item
$z_p \ =\ 1,\ z_q \ =\ 0,\ z_r \ =\ 1,\ u^j \ =\ 0,\ v^j \ =\ 0,\  
w^j \ =\ 0,\  t^j \ =\  1,\  x^j \ =\ 1;\  $
\item
$z_p \ =\  0,\  z_q \ =\  1,\  z_r \ =\  1,\  u^j \ =\  1,\ v^j \ =\ 0,\ 
w^j \ =\ 0,\  t^j \ =\  1,\ x^j \ =\ 1;\ $ and 
\item
$z_p \ =\ 1,\ z_q \ =\ 1,\  z_r \ =\ 1,\  u^j \ =\ 0,\ v^j \ =\ 0,\  
w^j \ =\ 1,\  t^j \ =\ 0,\ x^j \ =\ 1.$
\end{remunerate}
When $ c_j =  (z_p +  z_q )$, we need to verify that the
only exactly-one assignments of $c_j' $ are the following:
\begin{remunerate}
\item
$ z_p \ =\ 1,\ z_q  = 0,\  
u^j \ =\ 0,\ v^j \ =\ 0,\  
w^j \ =\ 0,\  
t^j \ =\ 1,\  x^j \ =\ 0,\  
c^j \ =\ 0,\ 
d^j \ =\ 0,$\\ 
$e^j \ =\  1,\  f^j \ =\ 0;$
\item
$z_p \ =\ 0,\ z_q \ =\ 1,\ u^j \ =\ 1,\  v^j \ =\ 0,\  w^j \ =\ 0,\ 
t^j \ =\ 1,\ x^j \ =\ 0,\  c^j \ =\ 0,$\\ 
$d^j\ =\ 0,\  e^j\ =\ 1,\  
f^j \ =\ 0;\  and\ $
\item
$z_p \ =\ 1,\ z_q \ =\ 1,\ u^j \ =\ 0, \ 
v^j \ =\ 0,\ w^j \ =\ 1,\  t^j \ =\ 0,\  
x^j \ =\ 0,\  c^j \ =\ 0,$\\
$d^j \ =\ 0,\  e^j \ =\  1,\  f^j \ =\ 0.$
\end{remunerate}
\qquad \end{proof}

\noindent
{\sc 1-Ex3Sat} $\rightarrow$ {\sc 1-Ex3MonoSat:}

Let $f$ be an instance of {\sc 1-Ex3Sat}.
Let $\displaystyle{ f = \bigwedge_{j=1}^m c_j}$. 
For each $c_j$ construct $c_j'$ as
follows:
Replace each negated literal  of the form $\overline{z_p}$ appearing in
the clause $c_j$ by a distinct new variable $y^j_p$ in $c_j$,
then add the clauses
$(z_p + y^j_p + a^j_p) \wedge (a^j_p + d^j_p + e^j_p) \wedge$
$ (a^j_p + f^j_p + e^j_p) \wedge (d^j_p + f^j_p + e^j_p)$. 
Note that for each negated literal, we introduce new copies of the 
auxiliary variables $a^j_p, \ldots, f^j_p$.
See Figure \ref{1-Ex3-1-3monofig3.fig} for an example.

Let $\displaystyle{f' = \bigwedge_{j=1}^m c_j'}$. 
Then $f'$ is an instance of {\sc 1-Ex3MonoSat} obtained from $f$.
The result follows from
Lemma \ref{ex-one} and the fact that for all variables
 $x$ and $y$, $(x+y)$ is exactly-one-satisfiable {\it if and only if}
$x= \overline{y}$. 

\begin{figure}[tbp]
\centerline{\epsffile{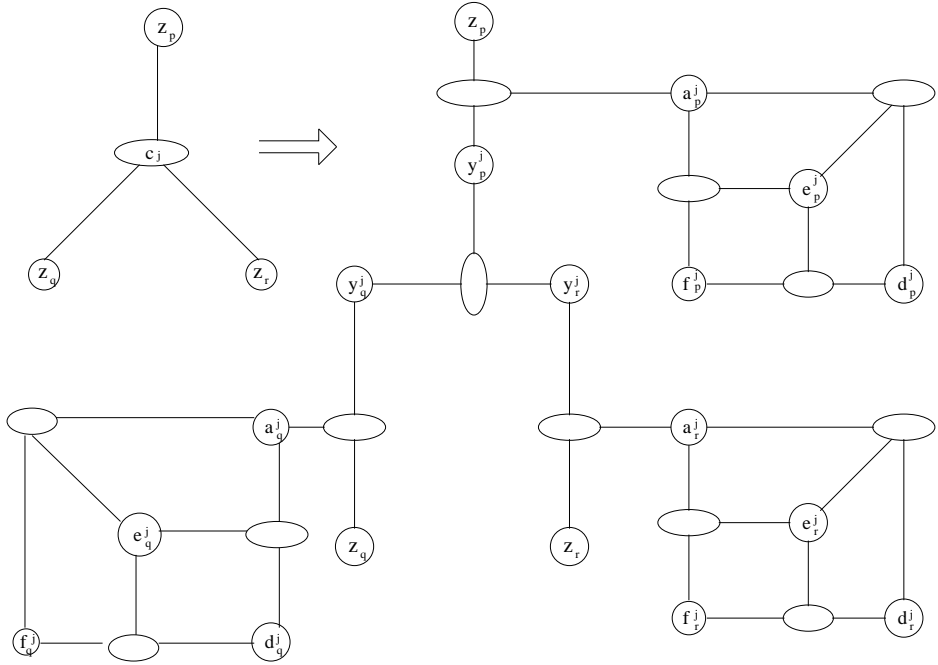}}
\caption{Figure illustrating the reduction from 
{\sc 1-Ex3Sat} to {\sc 1-Ex3MonoSat}. 
The figure illustrates the construction for a 3 literal clause $c_j$ 
which contains all negated literals.}
\label{1-Ex3-1-3monofig3.fig}
\end{figure}

\begin{figure}[tbp]
\centerline{\epsffile{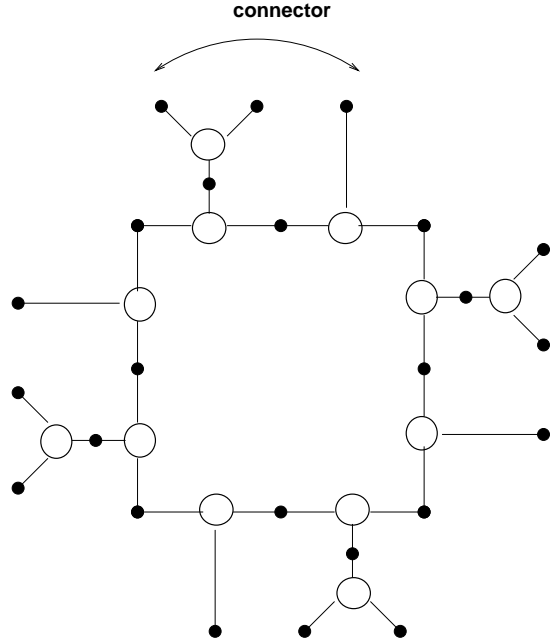}}
\caption{ Variable Configuration for reduction to {\sc X3C}. The black
dots represent element nodes while the ellipses denote the triples.}
\label{x3c1.fig}
\end{figure}

\begin{figure}[tbp]
\centerline{\epsffile{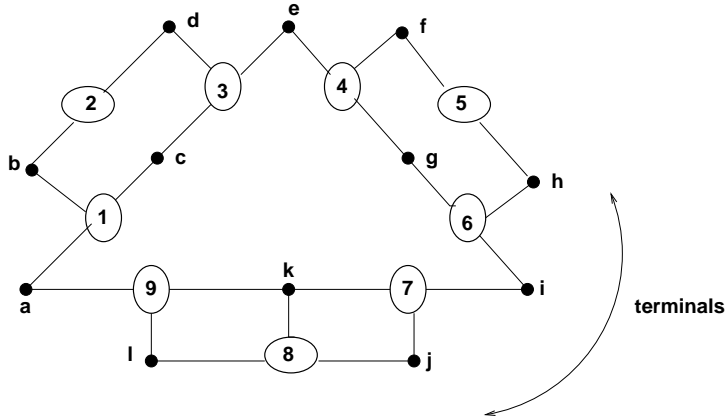}}
\caption{Figure illustrating the clause configuration. The sets 
$\{l, a, b \}, \{d, c, f \}, \{j, i, h \}$ represent the three terminals.
The vertices labeled $c, g, k$ are the internal elements.}
\label{x3c3.fig}
\end{figure}

\begin{figure}[tbp]
\centerline{\epsffile{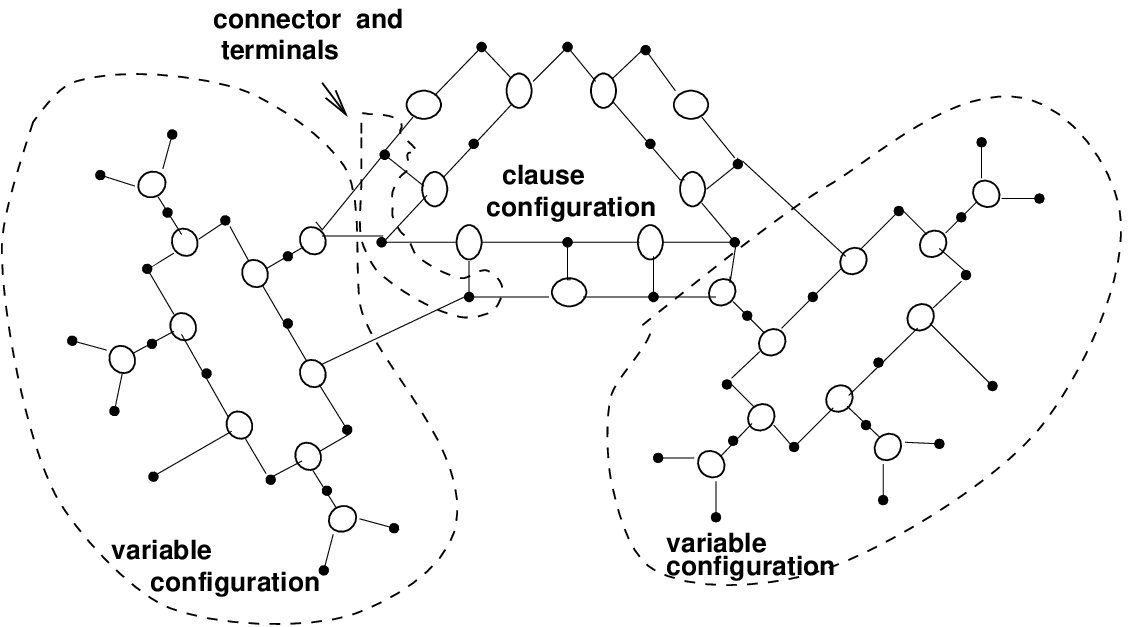}}
\caption{ Figure illustrating the way clause and the variable
configurations are attached.}
\label{x3c4.fig}
\end{figure}

\noindent
{\sc 1-Ex3MonoSat} $\rightarrow$ {\sc X3C:}

\noindent
Although Dyer and Frieze \cite{DF86} do not observe this, the reduction
given in their paper \cite{DF86} from {\sc 1-Ex3Sat} to 
{\sc X3C} is actually parsimonious. The reduction presented here is
essentially the same as given in \cite{DF86}, 
except that we start from an instance of {\sc 1-Ex3MonoSat}. 
Thus in our reduction we do not have to take care
of negated literals. We now describe the reduction.
Each variable is represented by  a cycle of 
3 element sets. If the variable occurs $r$ times in the {\sc 1-Ex3Sat} 
instance then there are $2r$ sets, with each successive
pair of sets sharing an element.
This cycle is augmented with $r$ additional sets and 
$2r$ elements by adding a 3-set to one of the external
elements in each pair. The 3 elements now corresponding 
to an appearance of $v_j$ will be called a connector.
The variable $v_j$ is set to true if and only if all three
connector elements are covered by the cycle when $v_j$
appears in the corresponding clause.
Figure \ref{x3c1.fig} illustrates the variable component.
Next, consider each clause $c_i$. Each $c_i$ is
represented by a configuration shown in Figure~\ref{x3c3.fig}.
This has 12 elements and 9 sets. Of the 12
elements, 3 are internal and the rest are grouped in groups of 3. 
Each group of 3 elements is called a terminal of $c_i$.
Finally, we connect a  clause component to the variable component as follows.
For each $v_j \in c_i$ we identify  three distinct  connector elements with 
one of the terminals in $c_i$. The construction is depicted in Figure
\ref{x3c4.fig}.
Let $G$ denote the graph obtained as result
of the construction. Planarity of $G$ follows by the fact that each 
component replacing a variable and a clause is planar and the components
are joined in a planarity preserving way.  
We first prove that there is an exact cover of $c_i$ configuration if and
only if exactly one terminal is covered externally, when we restrict the
covering such that either none or all three of the elements in each terminal
are covered externally.
But the configuration has the property that 
each of the three internal elements appear in 3
of the 9 sets and no two appear in the same set.
It follows that if this configuration forms part of an
exact cover by 3-sets, then exactly three of the sets must be used,
hence nine of the twelve elements will be covered internally. 
Moreover this can only be done 
so that exactly one of the terminals will be left uncovered.
This uncovered terminal is covered by sets in the variable
configuration and 
amounts to setting the literal true. 
It can then be argued that the exact cover by 3 sets
has a solution if and only if the corresponding {\sc 1-Ex3MonoSat} instance
is satisfiable.
It is easily verified  that the reduction is 
parsimonious. This is because the clause configuration
forces precisely one variable to be set to true and the
other two literals to be false. Moreover, for each 
satisfying assignment of {\sc 1-Ex3MonoSat}, there is exactly
one way the sets can be chosen so as to have an exact
cover.  Hence, the reduction is parsimonious. \qquad \end{proof}

\begin{corollary}
The problems  {\sc Pl-Ex3Sat, Pl-1-Ex3Sat} and\\
{\sc Pl-1-Ex3MonoSat}  are 
{\sf NP}-complete. 
The problems {\sc \#Pl-Ex3Sat, \#Pl-1-3Sat, \#Pl-1-Ex3Sat,
\#Pl-1-Ex3MonoSat} and {\sc \#Pl-X3C} are {\sf \#P}-complete.
\end{corollary}

\section{Planar graph  problems}\label{sec:graph}
\subsection{Overview of our proofs:}
In this section we give parsimonious/weakly parsimonious and
planarity preserving
reductions from {\sc Pl-1-Ex3MonoSat} to various graph problems. The 
problems considered here are {\sc Minimum Vertex Cover,
Minimum Dominating Set, Clique Cover,
Feedback Vertex set, Partition into Claws, Partition in Triangles}
and {\sc Bipartite Dominating Set}. Previously, reductions
showing that these problems were {\sf NP}-hard frequently did not
preserve the number of solutions. 
Central to the proofs is the  reduction (called $RED1$) 
from  {\sc 1-Ex3MonoSat}
to {\sc 3Sat} with the property that every formula is 
mapped to a formula in which
all satisfying assignments satisfy exactly one literal in each three
literal clause. This in turn enables us to obtain (weakly) parsimonious 
reductions from {\sc Ex1-3MonoSat} to the problems considered here.

\subsection{Reduction RED1}\label{sec:red1}
\noindent
{\bf RED1:}
Let $RED1$ be a mapping from an instance 
$\displaystyle{f = \bigwedge_{j=1}^{m}c_j}$ of {\sc 1-Ex3MonoSAT}
to an instance $\displaystyle{f' = \bigwedge_{j=1}^{m}c_j'}$ of {\sc 3Sat},
where for $c_j =  (x + y + z)$, 
$c_j' = (x + y + z) \wedge (\overline{x} + \overline{y})$
$ \wedge (\overline{x} + \overline{z})$
$ \wedge (\overline{z} + \overline{y})$

\begin{figure}[tbp]
\centerline{\epsffile{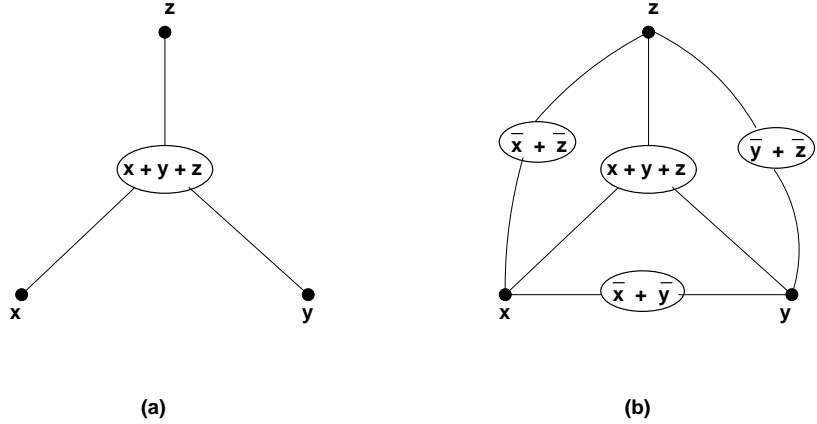}}
\caption{Figure illustrating the reduction $RED1$ discussed in \S\ref{sec:red1}.
Observe that the reduction is planarity preserving.}
\label{red3.fig}
\end{figure}

\begin{lemma}
The formula $f'$ has the following properties: \\
(1) The satisfying assignments of $f'$ are {\bf exactly} the
exactly-one satisfying assignments of $f$. \\
(2) In any satisfying assignment of $f'$, all but one clause in $c_j'$
is exactly-one satisfiable.\\
(3) Each variable in the formula $f'$ occurs at least twice negated and 
at least once unnegated.\\
(4) RED1 is  planarity preserving.(See Figure \ref{red3.fig}.)\\
(5) RED1 is parsimonious.
\end{lemma}
\begin{proof} By inspection. \qquad \end{proof}

We call each of 
the $c_j'$  a {\it clause group}. 
Observe that each $c_j'$ has four clauses,
one of which is a three literal clause and the others are two literal
clauses. 

\subsection{Weakly parsimonious reductions and basic graph problems}
\begin{theorem}\label{th:mvc}
There exists a planarity preserving and
weakly parsimonious reduction from {\sc 1-Ex3MonoSat} to each of the following
problems:  
(1) {\sc Minimum Vertex Cover}, 
(2) {\sc Minimum Dominating Set,}  
(3) {\sc Minimum Feedback Vertex Set,} and 
(4) {\sc Subgraph Isomorphism}. 
\end{theorem}
\begin{proof}

\noindent
(1) {\sc Minimum Vertex Cover:}

The reduction is from {\sc 1-Ex3MonoSat} and is similar to the one given
in \cite{GJ79} for proving {\sf NP}-hardness of 
{\sc Minimum Vertex Cover}.
Let $f$ be a {\sc 1-Ex3MonoSat} formula.
Apply $RED1$ to $f$ to obtain $f'$. 
Next, starting from $f'$ construct an instance $G(V,E)$
of the vertex cover problem as shown in Figure~\ref{figvc.fig} as follows: 
\begin{remunerate}
\item
Consider a clause group 
$c_j' = (x + y + z) \wedge (\overline{x} + \overline{y})$
$ \wedge (\overline{x} + \overline{z})$
$ \wedge (\overline{z} + \overline{y}).$
Corresponding to the  clause 
$(x+y+z)$, construct a triangle with vertices
$\{x, y, z\}$ and edges $\{(x, y),(x, z),(y, z) \}$. 
Corresponding to a clause 
of the form $(\overline{x} + \overline{y})$
add the edge  $\{(x,y) \}$. Call this the clause graph.

\item
For each variable $x$ that appears $i$ times we construct a simple cycle 
with $2i$ vertices. Let the odd numbered variables represent the negated
occurrences and the even numbered variables represent 
the unnegated occurrences. Call this the variable graph.

\item
Join the vertices of the clause graph to the vertices of the variable
graph as follows: 
Consider a clause group $c_j'$. Corresponding to 
a clause, join the triangle vertices $x, y$ and $z$ 
to the corresponding unnegated occurrences of $x,y,$ and $z$ in the cycles.
Corresponding to a clause of the form $(\overline{x}+\overline{y})$,
join the two vertices $x, y$
to the negated occurrence of the variables $x$ and  $y$ respectively.
Repeat the procedure for each clause group.

\end{remunerate}

\begin{figure}[tbp]
\centerline{\epsffile{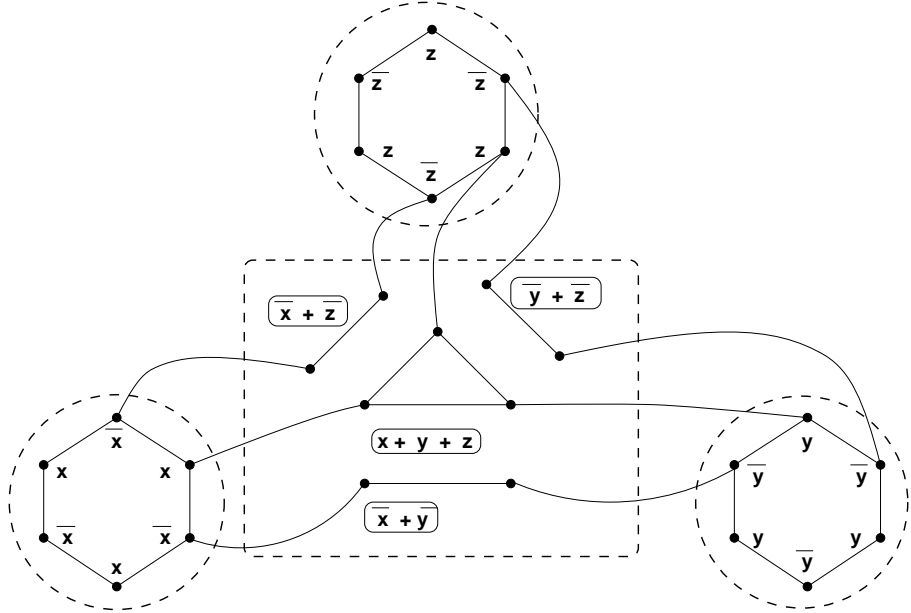}}
\caption{Figure illustrating the reduction from {\sc 1-Ex3MonoSat} 
to {\sc Vertex Cover}. The figure illustrating transformation 
of a clause group $(x + y + z)  \wedge (\overline{x} + \overline{y})
\wedge (\overline{x} + \overline{z}) \wedge (\overline{z} + \overline{y}).$
The dotted enclosures depict how to locally transform the clauses as well
as the variables so as to preserve planarity of the resulting graph.}
\label{figvc.fig}
\end{figure}

Now set $K = 1/2\sum C_i  + 2m + 3m$, where
$C_i$ is the length of the cycle of the variable $i$, and
$m$ is the number of clause groups in $f'$. The reduction is illustrated in
Figure \ref{figvc.fig}.
\begin{claim}
(1) The formula $f'$ is satisfiable if and only if the graph $G$
has a vertex cover of size $K$.\\
(2) The reduction is planarity-preserving and
weakly parsimonious. 
\end{claim}

\begin{proof}
\noindent
{\sf Part 1:}
Observe that for any vertex cover
one needs to pick at least half of the nodes from each cycle,  two
of the three nodes from each triangle, and one from the simple edge
for each two-literal clause. 
(Recall that there are $3m$ two-literal clauses in $f'$.)
The sum is exactly $K$. 
Given this observation, the proof is similar to the one given in \cite{GJ79}.

\noindent
{\sf Part 2:}
It can be easily verified by observing Figure \ref{figvc.fig} that the
reduction is planarity preserving. 
To see that for each distinct satisfying assignment of $f'$, there are
$2^m$ distinct vertex covers of size $K$, one observes that
for each satisfying assignment, all but one clause in each clause group
has only one true literal. This forces the choice of vertices from the
clause graph, for all clauses having only one true literal.
For each satisfying assignment and for
each clause group there is one clause in the clause group in which both 
the literals are  true. For each such clause any of the two
vertices can be included in the vertex cover. Since there are
$m$ clause groups, we have $m$ such clauses and hence 
we have $2^m$ different vertex covers for each satisfying assignment.
\qquad \end{proof}

Note that our reduction shows that counting the number of vertex covers
of size $\leq k$ is {\sf \#P}-hard even if there are {\bf no} vertex covers of
size strictly less than $k$. For the next two results, we use such an
instance of \#-vertex cover for our reductions.

\noindent
(2){\sc Minimum Dominating Set:}

The reduction is  from the {\sc Minimum Vertex Cover} problem.
The reduction in \cite{Ka72} from {\sc Vertex Cover} to 
{\sc Dominating Set} can be 
easily modified to get a parsimonious reduction. 
Let $G_1 = (V_1, E_1)$, $V_1 = \{ v_1, \ldots v_n \}$ be an instance of the
{\sc Minimum Vertex Cover} problem. We construct an instance 
$G_2 = (V_2, E_2)$ of the {\sc Minimum Dominating Set} problem as follows.
There is one vertex in $V_2$ corresponding to every vertex in $V_1$. For each
edge in $G_1$ we also introduce two additional vertices and join them to the
two endpoints of the original edge.
Formally, $V_2 = U_1 \cup U_2$,  and $E_2 = A_1 \cup A_2$, where
\[U_1 = \{ u_i ~ | ~ v_i \in V_1 \} \]
\[U_2 = \{x_1^{ij}, x_2^{ij} ~ | ~ (v_i, v_j) \in E_1 \} \]
\[ A_1 = \{ (u_i, u_j) ~ | ~ (v_i, v_j) \in E_1 \} \]
\[ A_2 =  \{(u_i, x^{ij}_1), (x^{ij}_1, u_j), 
(u_i, x^{ij}_2), (x^{ij}_2, u_j) ~ | ~ (v_i, v_j) \in E_1 \}\] 

\begin{claim}
$G_1$ has  a minimum vertex cover of size $k$ if and only if $G_2$
has a minimum dominating set of size $k$.
Furthermore, the reduction  is planarity-preserving and parsimonious.
\end{claim}

\begin{proof}
It is easy to see that the reduction is planarity preserving.  
Consider a minimum  vertex cover 
$VC = \{v_{i_1}, v_{i_2}, \ldots, v_{i_k} \}$ of $G_1$. 
Corresponding to 
$VC$ we claim that there is exactly one dominating set in the graph $G_2$,
namely the vertex set  $DS = \{u_{i_1}, u_{i_2}, \ldots, u_{i_k} \}$. 
First note that 
for each edge in the original graph $G$ we have 4 new edges and 2 new vertices
in $G_2$.
Consider a pair of nodes of the form $x^{ij}_1, x^{ij}_2$ 
connected to the nodes $u_i$ and $u_j$. It is clear that
the only way to dominate both $x^{ij}_1, x^{ij}_2$ by using only one node is
to include one of $u_i$ or $u_j$ in the dominating set.
We need to consider two cases.
First consider the case when exactly one of $v_i$ or $v_j$ is in $VC$.
Then it is clear that there is exactly one dominating set $DS$ in $G_2$ 
corresponding to $VC$. When both $v_i$ and $v_j$ are in $VC$, 
the minimality of $VC$
implies that at least one edge incident on $v_i$ 
and at least one edge incident 
on $v_j$ are covered solely by $v_i$ and $v_j$ respectively. 
This implies that both $u_i$  and $u_j$ have to be in any feasible 
dominating set of $G_2$.
Thus we have exactly one dominating set $DS$ in $G_2$ 
corresponding  to the vertex cover $VC$ in $G_1$.
Conversely, consider a minimum 
dominating set $DS = \{u_{i_1}, u_{i_2}, \ldots, u_{i_k} \}$ 
of size $k$ in $G_2$.
Consider an edge $(v_i,v_j)$ in $G$. 
If $u_i$ and  $u_j$ are not in $DS$, then by construction of $G_2$, both 
$x_1^{ij}$ and $x_2^{ij}$ are in $DS$. But we could then construct a new
dominating set $DS'$ of $G_2$ where
\[ DS' = DS - \{x_1^{ij}, x_2^{ij} \} \cup \{u_i \}. \]
Clearly $|DS'| < |DS|$ which is a contradiction to the assumption that
$DS$ is a minimum dominating set. Thus, $DS$ does not contain any 
vertex from the set $U_2$. We now claim that 
$VC = \{v_{i_1}, v_{i_2}, \ldots, v_{i_k} \}$ is a vertex cover of $G$. 
The claim follows by observing that corresponding to each edge 
$(v_i, v_j) \in E_1$,
at least one of the vertices $u_i, u_j, x_1^{ij}, x_2^{ij}$ are  in the set 
$DS$. We have already argued that $x_1^{ij}, x_2^{ij} \not\in DS$. Thus 
one of $u_i, u_j$ is in $DS$. The corresponding vertex in $VC$ is seen to cover
the edge $(v_i, v_j)$.  \qquad \end{proof}

\noindent
(3){\sc Feedback Vertex Set:}

The reduction is from {\sc Minimum Vertex Cover} problem. 
Starting from an instance $G_1(V_1, E_1)$ of the 
{\sc Minimum Vertex Cover} problem, we construct the graph $G_2$ that is
identical to the one given for the 
{\sc Minimum Dominating Set} problem.
By arguments similar to those 
given in the proof of {\sc Minimum Dominating Set} problem, it is easy to that
the $G_1$ has a minimum vertex set of size $K$ if and only if
$G_2$ has a feedback vertex set of size $K$ and that the reduction is 
parsimonious.

\noindent
(4){ \sc Subgraph Isomorphism:}

Follows directly  by taking the 
graph $H$ to be a simple cycle on $n$ nodes,
and the weakly parsimonious reduction from
{\sc 3Sat} to the {\sc Hamiltonian Circuit} problem given in
Provan \cite{SP86}. \qquad \end{proof}

\begin{corollary}
The problems
{\sc \#Pl-Minimum Vertex Cover, 
\#Pl-Minimum Dominating Set 
\#Pl-Minimum Feedback Vertex Set} and
{\sc \#Pl-Subgraph Isomorphism}
are {\sf \#P}-complete.
\end{corollary}

\subsection{Parsimonious reductions and other graph problems}
In this section we briefly discuss why the reductions studied by
\cite{GJS76,DL82} and \cite{DF86,DF85} from
{\sc X3C} to various other graph problems are parsimonious and planarity
preserving.
\begin{theorem}
There exist planarity-preserving  and parsimonious reductions 
from {\sc X3C} to each of the problems
(1) {\sc Minimum Clique Cover,}
(2) {\sc Partition into Claws,} 
(3) {\sc Bipartite Dominating Set,}
(4) {\sc Partition into Triangles} and
(5) {\sc Minimum Hitting Set}.
\end{theorem}

\begin{proof}
\noindent
(1) {\sc  Minimum Clique Cover:}

The reduction is the same as given in \cite{DF86,GJ79}.
Given an instance $I(X,C)$ of
{\sc X3C}  such that $|X| = 3p$ and $|C| = m$, we
construct an instance $G$  of the {\sc  Minimum Clique Cover} problem 
such that $G$ has a clique
cover with cliques of size 3 if and only if  $I$ has a solution. 
The reduction consists of 
replacing each triple in the instance of $I$ by a triangle and by replacing 
an edge from a triple to an element by a set of triangles. The reduction 
is illustrated in Figure \ref{cliquecover.fig}. 
Formally, for each element, we have a vertex in $G$.
Corresponding to each triple $t_i = \{ x_i, y_i, z_i \}$, 
and the associated edges $(t_i, x_i), (t_i, y_i), (t_i, z_i)$ we create the
subgraph as shown in Figure \ref{cliquecover.fig}. 
The graph $G$ obtained by carrying out the above reduction for each triple 
has $3p + + 9m$ vertices  and $18m$ edges.
The reduction is planarity preserving as each component is planar and they
are joined in a planarity preserving way.
We claim $I$ has a solution if and only if 
$G$ has a clique cover of size ($p + 3m$).  In particular as shown in 
\cite{DF86,GJ79}, if $t_1, \ldots t_p$ are the set of triples in an 
exact cover then the corresponding clique cover is constructed by taking
\[ \{\alpha^i, \beta^i, x_i \}, \{ \gamma^i, \delta^i, y_i \}, 
\{ \kappa^i, \pi^i, z_i \}   \{ t_1^i, t_2^i, t_3^i \} \]
whenever $ t_i = \{ x_i, y_i, z_i \}$ is in the exact cover and 
by taking the cliques
\[ \{\alpha^i, \beta^i, t_1^i \}, \{ \gamma^i, \delta^i, t^i_2 \}, 
\{ \kappa^i, \pi^i, t_3^i \} \]
when the corresponding triple $t_i$ is not in the exact cover.
Conversely, since $G$ has $3p + 9m$ vertices, 
if $G$ has a clique cover of size $p + 3m$ it implies that each 
clique consists of exactly 3 vertices. 
(Recall we do not have cliques of size four in  $G$.) 
The corresponding exact cover is given by choosing those $t_i \in C$
such that the triangles $t_1^i, t_2^i, t_3^i$ are in the clique cover.
Finally, we prove that the reduction is parsimonious. 
First note that if the triple triangle $\{t_1^i, t_2^i, t_3^i \}$
is not chosen then we have to choose the triangles 
$ \{\alpha^i, \beta^i, t_1^i \}, \{ \gamma^i, \delta^i, t^i_2 \}, 
\{ \kappa^i, \pi^i, t_3^i \}$ so as to cover the triple vertices.
Second, once the triangle 
$\{t_1^i, t_2^i, t_3^i \}$ is chosen there is exactly one way  for the 
auxiliary nodes  (and thus the element nodes) to be covered; 
namely the lower triangle corresponding to the covering
triple,  i.e. choosing  the triangles 
\[ \{\alpha^i, \beta^i, x_i \}, \{ \gamma^i, \delta^i, y_i \}, 
\{ \kappa^i, \pi^i, z_i \}.\]
These observations immediately imply that the reduction is parsimonious.

\begin{figure}[tbp]
\centerline{\epsffile{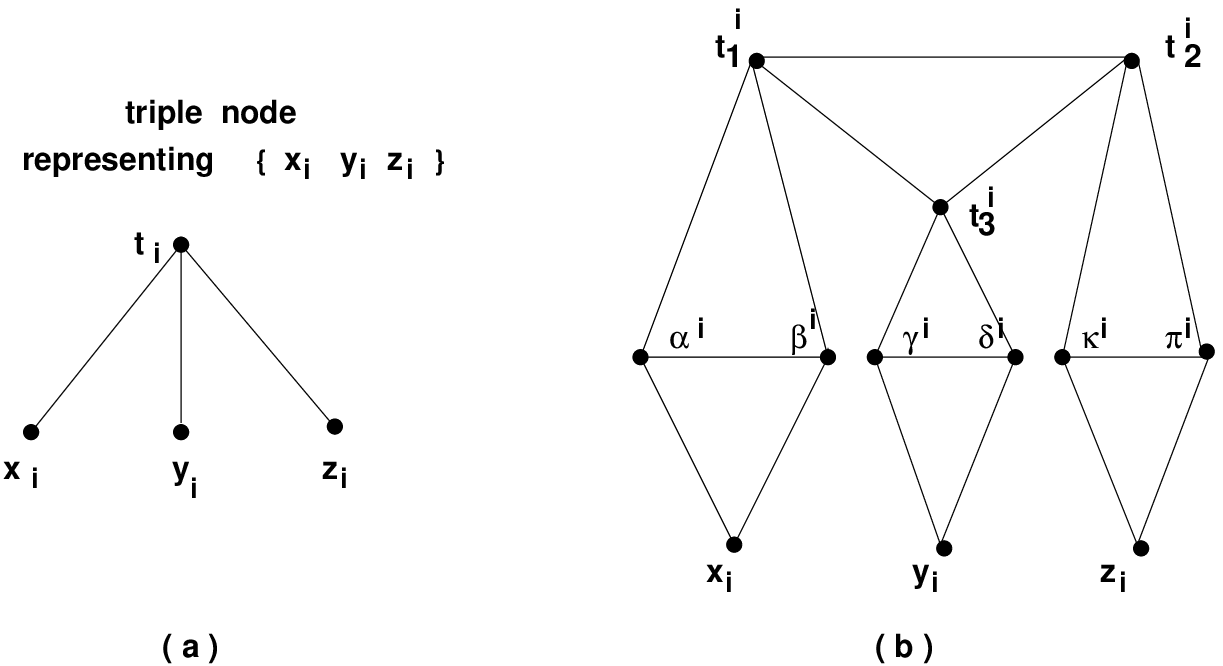}}
\caption{Figure illustrating the reduction from {\sc X3C} to 
{\sc Clique Cover}.}
\label{cliquecover.fig}
\end{figure}

\noindent
(2) {\sc Partition into Claws:}

The reduction is from {\sc X3C} and is the same as the one
given in \cite{DF86}. The reduction consists of the following steps.

\begin{remunerate}
\item
Construct the bipartite graph $G(C \cup X, E)$ corresponding to
the given instance $I(X, C)$ of {\sc X3C}.

\item
As in \cite{DF86}, we assume that each element vertex, 
appears in either two or three
sets, i.e. the element vertices  have a degree 2 or 3. 

\item
For each element of degree 3, we add an extra edge and for each 
element of degree 2, we add two extra edges. This is shown in 
Figure \ref{claws.fig}. Let $G_1$ denote the resulting graph.

\end{remunerate}

Clearly the reduction is planarity preserving.
We now recall the proof in \cite{DF86} to show that 
the edges of $G_1$ can be partitioned into a disjoint set of claws.
Note that each element vertex is adjacent to 
either 1 or 2  vertices of degree 1, it follows that each element node must
be the center of at least one claw. But each such element node has degree
4 and hence can be the center of exactly  one claw.
After removing the claws from $G_1$ the resulting graph $G_2$ has the property
that all element nodes have degree 1. This implies that the only way to
partition $G_2$ into claws is for each triple to have a degree of 0 or 3. 
Thus the triples with degree 3 induce a solution for the {\sc X3C} in an
obvious way. 
Conversely, given a solution for $I$, the above argument can be reversed to
yield a partition of the edges in $G_1$ into claws.
The following  
observations immediately imply that the reduction is parsimonious.
\begin{remunerate}
\item
there is a unique way to pick the claws in $G_1$ with the element nodes
as centers.

\item
In $G_2$,  each triple
vertex has degree 3 or 0 and each element node has degree 1 in $G_2$.
\end{remunerate}

\begin{figure}[tbp]
\centerline{\epsffile{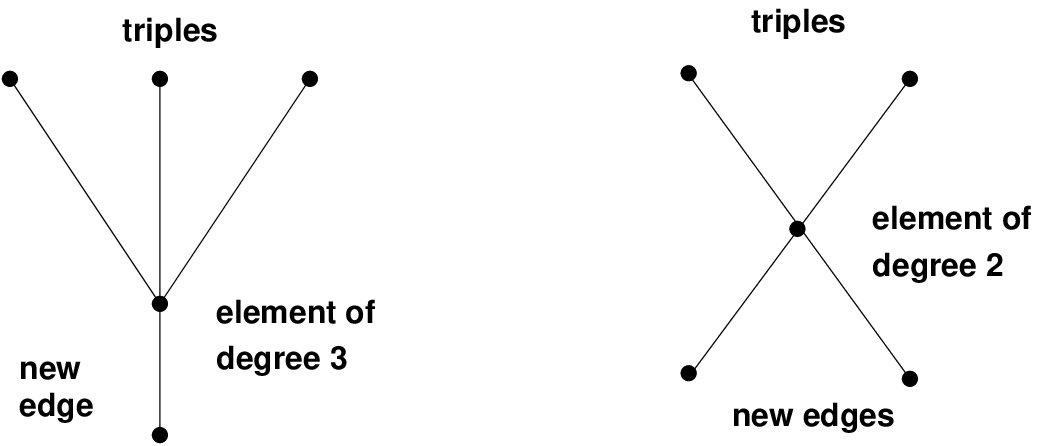}}
\caption{Figure illustrating the 
reduction from {\sc X3C} to {\sc Partition into Claws}.}
\label{claws.fig}
\end{figure}

\noindent
(3) {\sc Bipartite Dominating Set:}

Reduction from {\sc X3C}. 
The construction is similar to that in \cite{DF86}. Let
$I(X,C)$ be an instance of {\sc Pl-X3C} with each element occurring in at most
3 triples. We first construct 
the bipartite graph $G$ associated $I$. 
Next, we attach a 2-claw ($K_{1,2}$) 
to each triple vertex in $G$  as shown in Figure~\ref{dominating.fig}.  
(In \cite{DF86}, they add a path of length 2.)
Let $G'$ denote the graph obtained as a result of the transformation.
The construction is depicted in Figure~\ref{dominating.fig}. 
Since $G$ is bipartite and we added a claw as shown in 
Figure \ref{dominating.fig}, it follows that $G'$ is also bipartite.
Also note that the reduction is planarity preserving and thus $G'$ is planar.
Let the number of triples be $m$ and the
number of elements be $3p$. Then we set $k = p + m$. Now 
by arguments similar to those in \cite{DF86}, it is easy to see that $G'$
has a dominating set of size $k$ if and only if 
$I$ has a exact cover of size $p$. We prove the reduction is parsimonious. 
Consider a solution $S(I)$
for $I$. Since each triple in the solution covers three distinct element
nodes, these element nodes  can not be used to dominate the vertices  in $G'$
without increasing the cardinality of the solution for $G'$. This means that
for each of the $p$ triples chosen in the solution $S(I)$, we have exactly
one node in $G'$ that can be used in the dominating set so as to dominate all
the element nodes. Moreover due to the constraints on the size of the 
dominating set in $G'$, it follows that we can select exactly one vertex
per claw (the vertex with degree 3 and marked $a$) in the dominating set.
These observations imply that the reduction is  parsimonious.

\begin{figure}[tbp]
\centerline{\epsffile{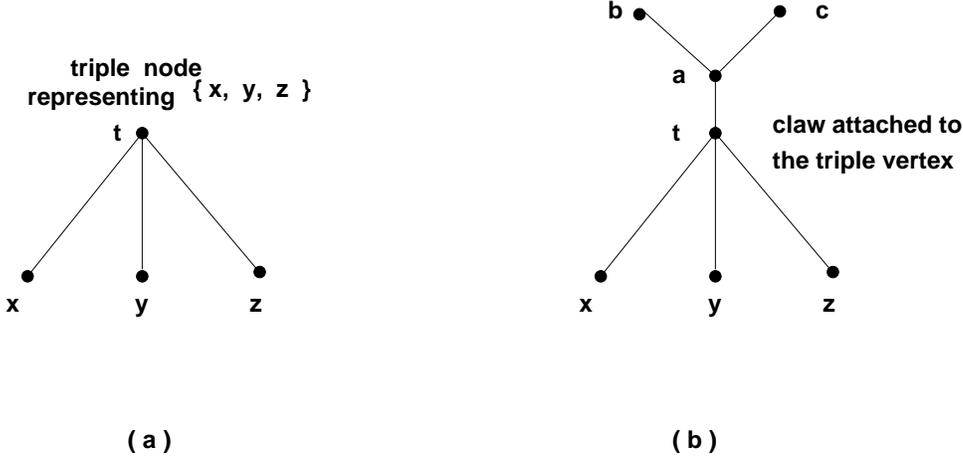}}
\caption{Figure illustrating the parsimonious reduction 
from {\sc X3C} to {\sc Bipartite Dominating Set}. It is easy to see that the
reduction preserves planarity of the graph.}
\label{dominating.fig}
\end{figure}

\noindent
(4) {\sc Partition into Triangles:}

The reduction is from {\sc X3C} and is the same as the reduction described in
the proof of {\sc Clique Cover}. 
Given that the resulting graph has no cliques
of size 4, the proof follows.

\noindent
(5){\sc Minimum Hitting Set:}

As given in  \cite{GJ79},  each instance of vertex cover can be seen to
be an instance of hitting set, in which every edge $(u,v)$ corresponds to
the set $\{u,v\}$. The elements of the set are simply the nodes of the graph.
The result now follows by noting that there is a 
weakly parsimonious reduction from {\sc 3Sat} to {\sc Minimum Vertex Cover}. 
\qquad \end{proof}

\begin{corollary}
The problems {\sc \#Pl-Minimum Clique Cover, 
\#Pl-Partition into Claws, \#Pl-Bipartite Dominating Set,
\#Pl-Partition into Triangles,}
 and {\sc \#Pl-Minimum Hitting Set} are {\sf \#P}-complete.
\end{corollary}
\begin{theorem}
Let $\Pi$ be one of the problems
in Table 1. It is {\sf \#P}-complete to count the number of solutions to
$\Pi$, even when one is given an instance of $\Pi$ and a solution which
is guaranteed to satisfy $\Pi$.
\end{theorem}
\begin{proof}
Starting from a 3CNF formula obtained in the proof of Theorem \ref{ambi},
we now do the same set of reductions discussed
in the earlier theorems to obtain an instance of the problem $\Pi$. Since
we know that the 3CNF formula is satisfiable, it follows that the instance
of $\Pi$ has a solution.  \qquad \end{proof}

\section{ Unique and  ambiguous  planar problems }\label{sec:unique}
Our parsimonious planar crossover box for {\sc 3Sat} 
can also be used to show that 
additional
problems for planar CNF formulas are {\it as hard as} 
the corresponding problems for arbitrary CNF formulas, 
with respect to polynomial time or
random polynomial time reducibilities. We briefly describe these results.
We first recall the definitions of {\sf D$^P$} and random polynomial
time reductions from \cite{Pa94,VV85}.

\begin{definition}\label{def5.1}
{\sf D$^P$} $= \{ L_1 - L_2 ~ | ~ L_1, L_2 \in {\sf NP} \}$. \\
Intuitively, a 
problem is in {\sf D$^P$} if it can be solved by asking
one question in {\sf NP} and one question in {\sf Co-NP}. 
\end{definition}

\begin{definition}\label{def5.2}
Problem $A$ is reducible to problem $B$ by a randomized
polynomial time reduction if there is a randomized
polynomial time Turing machine $T$ and 
a polynomial $p$ such that
\begin{remunerate}
\item
$\forall x ~  [x \not\in A \rightarrow T[x] \not\in B ].$

\item
$\forall x ~  [x \in A \rightarrow T[x] \in B$ with probability
at least $1/p(|x|) ]$. \QED
\end{remunerate}
\end{definition}

\begin{theorem}\label{unique-pl-3sat}
{\sc Unique-Pl-3Sat} is {\sf D$^P$}-complete
under randomized polynomial time reductions.
\end{theorem}

\begin{proof}
We modify the proof of the {\sf D$^P$}-completeness 
of {\sc Unique Sat} in \cite{VV85}, 
so that whenever their reduction outputs a boolean formula $f$,
output the planar formula  {\sc Pl}$(f)$  obtained by applying the 
parsimonious planar crossover box to $f$. The formula {\sc Pl}$(f)$ 
has exactly the same number
of satisfying assignments as $f$. 
In particular, {\sc Pl}$(f)$ is uniquely satisfiable
if and only if  $f$ is uniquely satisfiable. \qquad \end{proof}

A second example is the following :
\begin{theorem}\label{amb-pl-3sat}
{\sc Ambiguous-Pl-3Sat}  is {\sf NP}-complete.
\end{theorem}
\begin{proof}
Given an instance of an arbitrary 3CNF formula $f$, we first construct
a new formula using the same construction as in Step 1 of Theorem \ref{ambi}.
As pointed out in the proof of Theorem \ref{ambi}, the new formula is
ambiguously satisfiable if and only if the original formula is satisfiable.
We then do the same sequence of reductions as in Theorem~\ref{ambi} 
to obtain a planar formula that is ambiguously satisfied if and only if the 
original formula is satisfied. \qquad \end{proof}

Using the ideas similar to those in the proof of
Theorem \ref{1-valid}, we can prove that 
\begin{theorem}\label{amb-1-val-pl-3sat}
{\sc Ambiguous-1-Valid Pl-3Sat} is {\sf NP}-complete.
\end{theorem}

\begin{corollary}
{\sc Unique-1-Valid Pl-3Sat} is {\sf Co-NP}-complete.
\end{corollary}

\begin{proof}
To prove the membership in {\sf Co-NP}, consider an arbitrary formula\\
$F(x_1, \ldots, x_{n})$ which is an instance of {\sc 1-Valid Pl-3Sat}.
By the definition of {\sc 1-Valid} formulas an assignment {\bf v} to the
variables such that
{\bf v}[$x_1$] $ \ = \ $ {\bf v}[$x_2$] $ \ = \ldots $ 
{\bf v}[$x_n$] $ \ =  \ 1 $.
satisfies $F(x_1, \ldots, x_{n})$.
Now consider the formula 
$H(x_1, \ldots, x_{n}) = F(x_1, x_2, \ldots , x_n) \wedge
(\overline{x_1} \vee \overline{x_2} \ldots \overline{x_{n}})$.
$F$ is uniquely satisfiable if and only if $H$ is unsatisfiable. 
To prove {\sf Co-NP}-hardness,
given a formula $f(x_1, x_2, \ldots , x_n)$, we construct a formula $g$
such that 
\begin{center}
$g(x_1, \ldots, x_{n+1}) = 
[f(x_1, x_2, \ldots , x_n) \wedge x_{n+1}]
\bigvee
(\overline{x_1} \wedge \overline{x_2} \ldots \overline{x_{n+1}})$
\end{center}
Now using  ideas similar to those in the proof of
Theorem \ref{1-valid}, we obtain a planar formula $g_1$ with the
following properties:\\
(1) $g_1$ is 1-valid. \\
(2) $g_1$ is uniquely satisfiable {\it if and only if} $f$ is 
unsatisfiable. \qquad \end{proof}

Combining our parsimonious planar crossover box for {\sc 3Sat} and the
reductions to prove Theorem \ref{parsi}, we
get that exact analogues of Theorems~\ref{unique-pl-3sat}-
\ref{amb-1-val-pl-3sat} 
hold for each of the problems:
{\sc Ex3Sat,1-3Sat, 1-Ex3Sat} and {\sc 1-Ex3MonoSat}. 
Thus, we have the following corollary.

\begin{corollary}
Let $\Pi$ be one of the following problems:
{\sc Ex3Sat, 1-3Sat, 1-Ex3Sat} and {\sc 1-Ex3MonoSat}.
Then the problem {\sc Ambiguous-Pl-$\Pi$}
is {\sf NP}-complete  and the
problem {\sc Unique-Pl-$\Pi$} is {\sc D$^P$}-complete under randomized
polynomial time reductions. 
\end{corollary}

As a corollary of our parsimonious reductions, the unique versions
of many graph problems are also {\sf D$^P$}-complete.

\begin{corollary}
Let $\Pi$ be one of the problems   {\sc Pl-Partition into Triangles, 
Partition into Claws, Bipartite Dominating Set}.
Then the problem {\sc Ambiguous-$\Pi$}
is {\sf NP}-complete  and the problem 
{\sc Unique-$\Pi$} is {\sf D$^P$}-complete under randomized
polynomial reductions.
\end{corollary}

\begin{proof}
Given that each reduction in the  sequence of reductions
 {\sc 3Sat} $\rightarrow$ {\sc Pl-3Sat} $\rightarrow$ {\sc Pl-Ex1-3Sat} 
$\rightarrow$ {\sc Pl-X3C} is parsimonious,  
the fact that {\sc Unique 3Sat} is
{\sf D$^P$}-complete and the reduction from {\sc X3C} to each of the
problems mentioned above is parsimonious, the proof of the
corollary then is similar to the proof of Theorem~\ref{amb-pl-3sat}.
\qquad \end{proof}

\subsubsection{Non approximability results for integer linear programming}

Next, we give an application of our result that
{\sc Ambiguous Pl-3Sat} is {\sf NP}-complete and prove that it is not
possible to approximate the optimal value of the
objective function of a integer linear  program. 

An instance of integer linear program ({\sc ILP}) 
consists of a system of linear inequalities and
an objective function which is to be maximized(minimized); i.e.
Maximize(Minimize) 
{\bf c}$x$, subject to the constraints {\bf A}$x \leq $ {\bf b}.
The variables $x$ are allowed to take only integer values.
We say that a minimization problem $\Pi$ is $\epsilon$-approximable,  
$\epsilon > 1$, 
(or has an $\epsilon$-approximation) if there is a polynomial time algorithm
that given an instance $I \in \Pi$ finds a solution which is within a factor
$\epsilon$ of an optimal solution for $I$. 

\begin{theorem}\label{ILP}
Unless {\sf P} = {\sf NP}, 
given an instance of the problem {\sc ILP} and a feasible solution,
the maximum (minimum) value of the objective function 
is not polynomial time $\epsilon$-approximable 
for any $\epsilon > 1$, 
even when the bipartite graph associated with the set of 
constraints is planar.
\end{theorem}

\begin{proof}
We prove the theorem for the maximization version of the problem. 
The proof for the  minimization version is similar and hence omitted.

\noindent
{\bf Step 1:}
Given a {\sc 3Sat} formula $f(x_1, x_2, \ldots , x_n)$, we construct a formula
\begin{center}
$g(x_1, \ldots, x_{n+1}) = 
[f(x_1, x_2, \ldots , x_n) \wedge x_{n+1}] \vee
(\overline{x_1} \wedge \overline{x_2} \ldots \overline{x_{n+1}})$
\end{center}
It follows that for any assignment {\bf v}, 
{\bf v}[$g(x_1, \ldots, x_{n+1})$] $= 1$  if and only if either
(i) {\bf v}[$f(x_1, x_2, \ldots , x_n)$] $= 1$  and 
{\bf v}[$x_{n+1}] = 1 $, {\em or}
(ii) {\bf v}[$x_1$] $=$ {\bf v}[$x_2$] $= \ldots$ {\bf v}[$x_{n+1}$]  $= 0$.

\noindent
{\bf Step 2:}
Starting 
from $g(x_1, \ldots, x_{n+1})$, we  construct {\sc Pl-3Sat} formula\\
$\hat{g}(x_1, x_2, \ldots x_{n+1}, t_1, \ldots t_m)$ such that
$\hat{g}(x_1, x_2, \ldots x_{n+1}, t_1, \ldots t_m)$ is satisfiable
if and only if
$g(x_1, x_2, \ldots x_{n+1})$ is satisfiable.

The construction  can be carried out in a similar fashion as in Step 2 in proof
of Theorem \ref{ambi}. We therefore omit the details here.

\noindent
{\bf Step 3:}
Let $\hat{g}  = G_1 \wedge G_2 \ldots G_r$. Construct a new 
{\sc 1-Ex3-MonoSat} formula $h$ from $\hat{g}$ such that $\hat{g}$
is satisfiable if and only if $h$ is satisfiable.
Let $h  = C_1 \wedge C_2 \ldots C_p$. 
Replace each clause
$C_i = (x_{i_1} + x_{i_2} + x_{i_3})$ by the inequality
$ (x_{i_1} +  x_{i_2} + x_{i_3}) \geq 1$. All the inequalities
corresponding to the clauses make up the constraints.
We also add constraints that 
$\forall i, ~ x_i \in \{0,1 \}$. The 
objective function is now simply $x_{n+1}$.
It is easy to verify that
the maximum value of the objective function is exactly $1$
if  $f(x_1, x_2, \ldots , x_n)$
is  satisfiable and  is 0 otherwise. 
Hence it follows that unless {\sf P} = {\sf NP} the problem ILP 
has no polynomial time $\epsilon$-approximation algorithm 
for any $\epsilon > 1 $. \qquad \end{proof}

\section{Conclusions and open problems}\label{sec:conclusions}
We showed  that for many problems $\Pi$
studied in the literature, the problem {\sc \#Pl-$\Pi$, Ambiguous-Pl-$\Pi$}
and {\sc Unique-Pl-$\Pi$} are 
{\em as hard as} the respective problems
{\sc \#$\Pi$, Ambiguous-$\Pi$} and 
{\sc Unique-$\Pi$} with respect to polynomial time or
random polynomial time reducibilities. 
We note that the problem {\sc \#Pl-Hamiltonian-Cycle} was proved to be
{\sf \#P}-complete by Provan \cite{SP86}. 
We can give an alternate proof of the
{\sf \#P}-hardness of 
{\sc \#Pl-Hamiltonian-Cycle} by a reduction from a variant of $RED1$. 
The reduction is significantly
more complicated than that in \cite{SP86}. Consequently, we omit it here.

As corollaries of our results, we have shown that many planar problems 
are complete for the class {\sf NP}, {\sf \#P} and
{\sf D$^P$}. 
Our results and their proofs provide the following general
tools for proving hardness results for planar problems:
\begin{remunerate}

\item
We have shown how parsimonious and weakly parsimonious crossover boxes
can be used to prove the {\sf \#P}-hardness of many planar counting problems.
These ideas were used to prove the
{\sf \#P}-hardness of problem {\sc\#1-Valid Pl-3Sat}.

\item
We extended the class of basic planar CNF satisfiability problems
that are known to be {\sf NP}-complete.
Previously only {\sc Pl-3Sat} \cite{DL82} and {\sc Pl-1-Ex3Sat} \cite{DF86}
were known to be {\sf NP}-hard.
We expect that the variants of the problem
{\sc Pl-3Sat}  shown to be {\sf NP}-hard here will be useful
in proving hardness results for many additional
planar problems. 
In particular, we have already
shown that the problem {\sc Pl-1-3MonoSat}  and its variant {\em RED1} are
especially useful in proving the {\sf \#P}-hardness of
many planar graph problems.
\item
We have shown that the problem {\sc Ambiguous-Pl-3Sat} 
can be used to prove the
non-approximability of linear integer programming. 
Recently,
there has been a lot of research in the area of approximability
of graph and combinatorial problems and
the tools for showing negative results are few. 
Our proof of the non-approximability of the minimum or maximum
objective value of an integer linear 
program is direct and significantly different from 
the proof given in Kann \cite{Ka93} or Zuckerman \cite{Zu93}. 
Moreover, in \cite{HMS95}, 
we show how to use the {\sf NP}-completeness of 
{\sc Ambiguous-Pl-3Sat} to
show the non-approximability of several constrained optimization problems
even when restricted to {\em planar instances}. 
(The results in \cite{Ka93,Zu93} do not hold for planar instances.)
\end{remunerate}
Finally, the  results presented here  and their
proofs suggest a number of open problems including the following:
\begin{remunerate}
\item
Can natural planar problems be found that are complete for additional
complexity classes such as {\sf PSPACE, \#PSPACE, MAX SNP, MAX $\Pi_1$} etc ?
(In recent papers \cite{HMS94,HMS95,MH+95,HMR+94}, 
we partially answer this question by  showing
that a number of problems are complete for the classes 
{\sf PSPACE, \#PSPACE, MAX-SNP, MAX $\Pi_1$,}
even when restricted to planar instances.)
\item
Valiant \cite{Va79a} has shown that the problem {\sc \#2Sat} is 
{\sf \#P}-complete.
How hard is  the problem {\sc \#Pl-2Sat} ?
We conjecture that the problem is {\sf \#P}-complete,
but it seems to us that  different techniques 
than the ones used here are required to prove this.

\item
We have shown that many unique satisfiability problems are complete
for {\sf D$^P$},
even when restricted to planar instances. 
Using our parsimonious reductions, we then proved the
{\sf D$^P$}-completeness of a number of graph problems for planar graphs. 
A number of such problems for planar graphs remain open. For example,
how hard is the problem {\sc Unique-Pl-Hamiltonian Circuit} ?

\item
Do results similar to the ones proved in this paper hold for other restricted
classes of graphs, e.g.. intersection graphs of unit disks and squares?
Such  graphs have been studied extensively 
by  \cite{CCJ90,HM85,MS84} in
context of image processing, VLSI design, geometric location theory,
and network design.
\end{remunerate}

\vspace*{.2in}

\noindent
{\bf Acknowledgments:} 
We thank the anonymous referee for invaluable suggestions. These
suggestions significantly improved the quality of presentation 
and helped us in correcting a number of errors in the earlier draft.

\newpage 

\baselineskip = 0.9\normalbaselineskip

\end{document}